\documentclass[%
 reprint,
 amsmath,amssymb,
 aps,
]{revtex4-2}

\usepackage{graphicx}
\usepackage{dcolumn}
\usepackage{bm}

\usepackage{bbm}
\usepackage{macros}
\begin{document}


\title{ Criticality in 1-dimensional field theories with mesoscopic,
  \\  infinite
  range interactions  }

\author{$ ^{ab}$Kurt Langfeld}

\affiliation{
$^a$School CDMS,  Western Sydney University, 
Penrith NSW 2751, Australia 
}

\author{$^b$Amanda Turner}
\affiliation{
  $^b$University of Leeds, Leeds, LS2 9JT, UK 
}

\date{\today}

\begin{abstract}
This research investigates a novel class of
one-dimensional theories characterised by a distinctly defined
infinite interaction range. We propose that such theories emerge
naturally through a mesoscopic feedback mechanism. In this
proof-of-concept study, we examine Ising-type models and a model with
continuous O(3) symmetry, and demonstrate that the natural emergence 
of phase transitions, criticality, spontaneous symmetry breaking and
previously unidentified universality classes is evident.  The framework introduced here holds particular relevance for monolayer spintronics research, where the ultimate goal is to achieve a strong ferromagnetic order at room temperature. 
  
\end{abstract}

\maketitle

\section{Introduction }

Phase transitions and critical phenomena in low-dimensional systems
have long served as a testing ground for statistical field theory. The
Hohenberg--Mermin--Wagner
theorem~\cite{Mermin_wagner_1966,Hohenberg1967} provides one of the
cornerstones of this field, establishing that systems with continuous
symmetries and short-range interactions cannot exhibit spontaneous
symmetry breaking at finite temperature in one or two dimensions. The
theorem applies to models such as the isotropic Heisenberg or XY
ferromagnet, where the order parameter transforms continuously under
SU(2) or U(1) symmetry. However, the argument does not constrain
systems with discrete symmetries---such as the Ising model---or those
with infinite- or long-range couplings, where mean-field behaviour can
dominate even in one dimension. 

\medskip 
Dyson~\cite{Dyson1969}
demonstrated that phase transitions can occur in one-dimensional
Ising models with algebraically decaying couplings, $J(r) \sim
r^{-\alpha}$ for $1 < \alpha < 2$, and later rigorous work by
Fr\"ohlich and Spencer~\cite{Froehlich_spence_1982}  established a
transition for $1/r^2$ interactions. These studies clarified that the
presence or absence of long-range order depends not on dimensionality
alone, but crucially on the effective range and structure of
interactions. The existence of phase transitions  for 
the one-dimensional ferromagnetic Ising model with long-range two-body
interaction has subsequently been an object of much
research~\cite{Cassandro2009,Cassandro2014,Turban_2016,Halperin2019,
  Suzuki_2025}.

\begin{figure*}
\includegraphics[height=5cm]{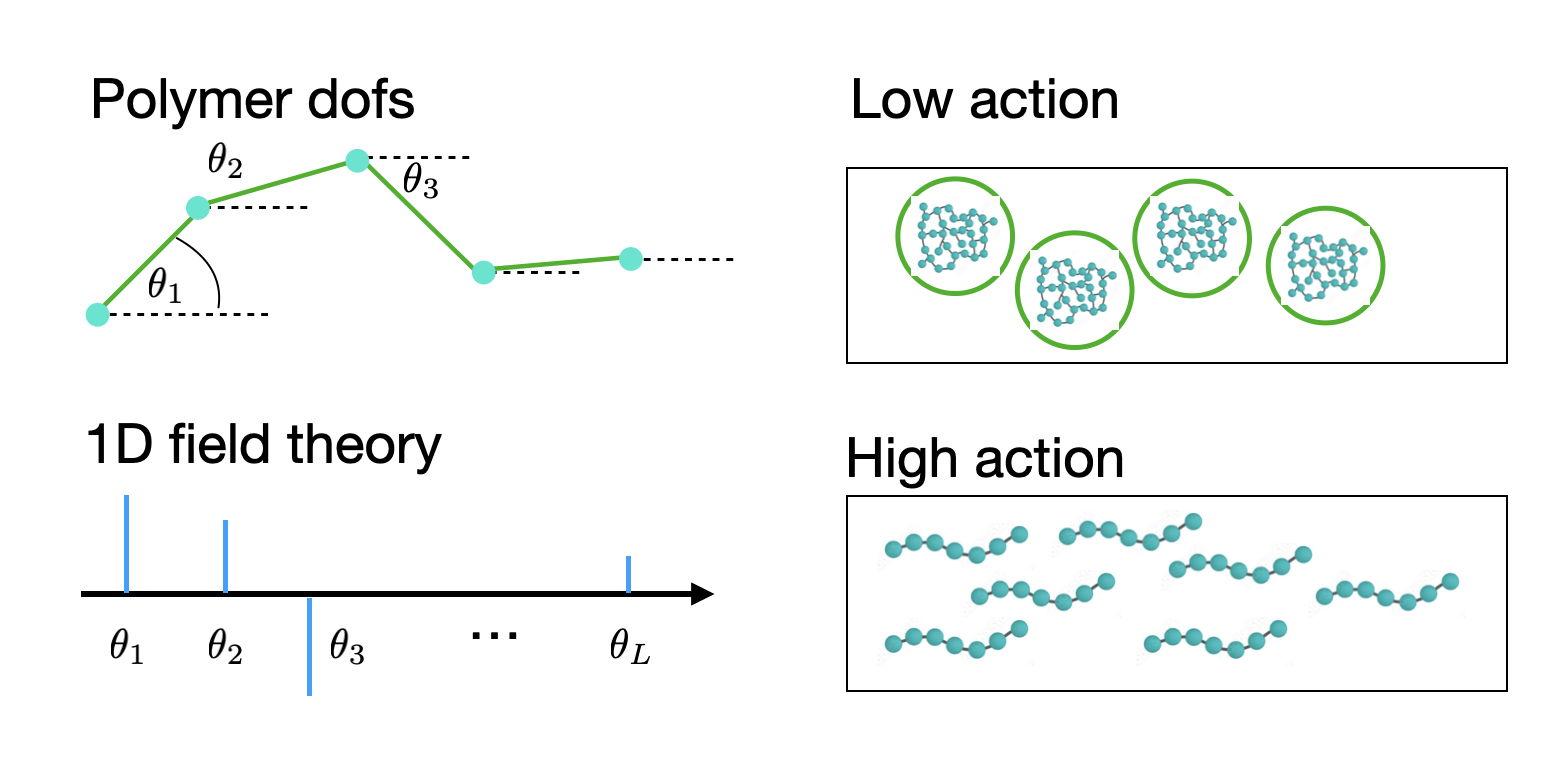}  \hspace{0.5cm}
\includegraphics[height=5cm]{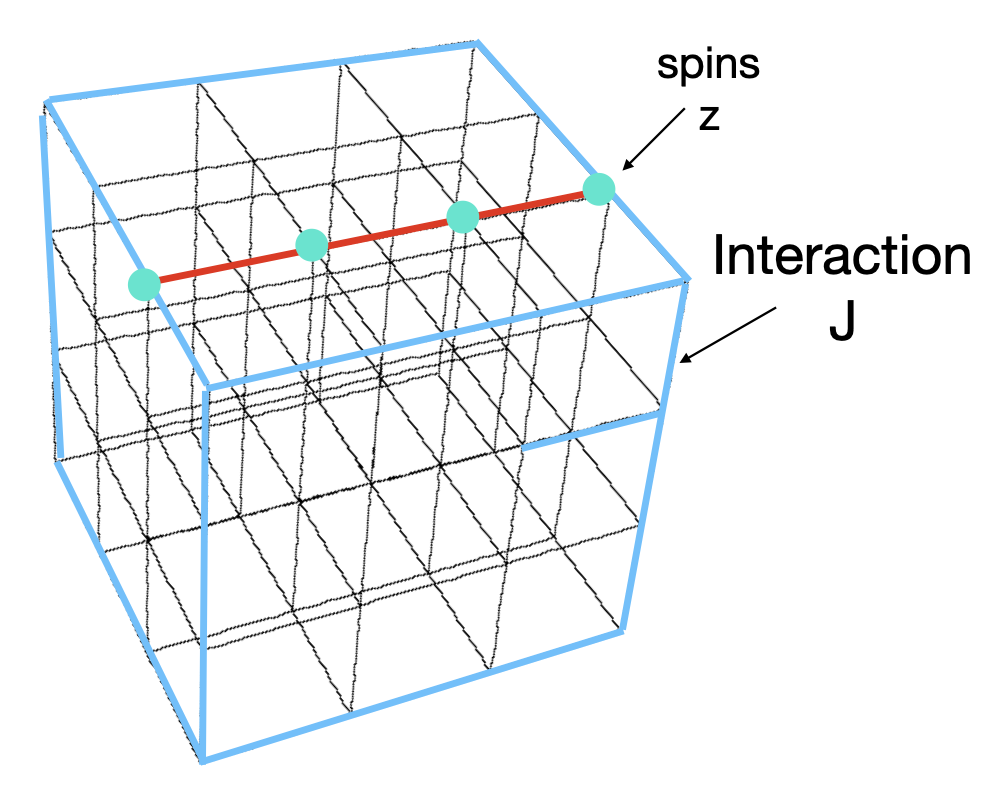}  \hspace{0.5cm}

  \caption{\label{fig:0} Left and Middle: Effective statistical theory for
    polymer thermodynamics with mesoscopic feedback on model
    parameter. Right: Effective 1D field theory emerging at the
    1-dimensional subdomain of a local 3D statistical field theory.  }
\end{figure*}

\medskip
In this paper, we develop novel one-dimensional systems in which the
interaction between elementary degrees of freedom depends on
mesoscopic observables such as the magnetisation or energy
density. This feedback mechanism endows the model with effectively
infinite-range interactions while preserving a clear physical
interpretation in terms of collective mesoscopic processes. The
resulting theory can be viewed as a generalisation of the Curie--Weiss
model~\cite{stanley1971introduction,Kochmanski_2013}, where each 
spin interacts equally with every other spin. The Curie--Weiss
model features an infinite range interaction, but because of the ``democratic" interaction of all spins with each other, it does not have a spatial structure to define dimensionality. Our models below, however, share the infinite range interaction with the Curie--Weiss model but inherit dimensionality, geometry, lattice structure, size and topology of the related model with local interactions. We will find that it retains analytic tractability but allowing for nontrivial extensions of criticality. 

\medskip
Our initial focus is on Ising-type models with discrete symmetry. We are interested in robust long range order and phase transitions. The interest in these topics has seen renewed interested with the discovery of Monolayer ferromagnets — atomically thin materials that exhibit ferromagnetism down to a single layer — which are considered as a major breakthrough for spintronics~\cite{Gong2017,Huang2017,Jenkins2022}. Transition temperatures are of particular interest for the quest for room temperature, mono-layer ferromagnets~\cite{Bonilla2018}. We will demonstrate that a mesoscopic feedback mechanism can produce stable long range order even in 1-dimensional systems.  
Within our 
framework, we derive closed-form expressions for the free energy and
demonstrate how the nature of the feedback term determines the order
of the transition. When the free energy depends quadratically on the
order parameter, the transition is continuous (second order); when
cubic or higher-order terms dominate, the transition becomes
discontinuous (first order). These results reproduce the expected
Landau-theoretic structure of phase transitions while revealing new
possibilities for feedback-induced universality classes in
one-dimensional systems.  

\medskip
The framework naturally extends to continuous symmetries. Motivated by
the question of whether the proposed mechanism can induce long-range
order in systems with continuous degrees of freedom, we examine the
one-dimensional $O(3)$ model under analogous mesoscopic
coupling. Despite the one-dimensional geometry, we find that the
feedback mechanism induces spontaneous symmetry breaking for
sufficiently strong coupling. This behaviour demonstrates that
mesoscopic feedback can effectively act as a nonlocal mediator of
interactions, producing ordered phases even where conventional
short-range models would remain disordered. The $O(3)$ case thus
serves as an important conceptual extension, linking the generalised
Landau-theoretic picture developed here to a broader class of
continuous-symmetry systems. 

\medskip
Beyond their specific analytic results, these models provide a
framework for exploring emergent universality in systems with
feedback-induced long-range correlations. Universality---the
insensitivity of critical behaviour to microscopic details---has long
been a central organising principle of statistical
mechanics~\cite{osti_7362153,Zinn-Justin2021,Sachdev_2011}. Its
manifestations range from phase transitions in magnets and superfluids
to collective behaviour in biological and neural
systems~\cite{Mora2011,Cavagna_Giardina_2014,Bryngelson87,NELSONONUCHIC200087}. The
density-of-states approach developed here provides a complementary
route to universality, in which critical scaling emerges not from
proximity in real space, but from a collective coupling across the
system. 

\medskip
The remainder of this paper is organised as follows. Section~\ref{sec:models} introduces the class of models and formulates the mesoscopic feedback mechanism. Section~\ref{sec:ising} presents analytical results for the generalised Ising model, including its free energy, order parameters, and critical exponents. Section~\ref{sec:o3} extends the analysis to the $O(3)$ model, demonstrating how feedback-driven coupling can produce symmetry breaking in continuous systems. Section~\ref{sec:conclusions} summarises our conclusions.

\subsection{Rise of 1D field theory with mesoscopic long-range interactions \label{sec:models}}

In this paper we introduce a novel class of one-dimensional non-local field
theories in which the fundamental constituents interact through
conventional local interactions, yet the parameters governing these
interactions are dynamically dependent on global field
configurations. 

\medskip 
The standard 
Ising or Heisenberg model have short-range interactions with no long range order in one dimension. In the Kac limit~\cite{Kac1963}, the interaction potential is rescaled as so that its range becomes large compared to the inter-particle distance but remains small compared to the system size.  One obtains a mean-field–type description while retaining the system’s underlying geometry. This construction provides a mesoscopic, finite-range analogue of infinite-range Curie–Weiss interactions. Our models described below have an infinite range interaction as the Curie–Weiss but retain the system’s underlying geometry and dimensionality as models in the Kac limit. 

\medskip 
Our mesoscopic feedback mechanism allows the system's
parameters—such as coupling strengths or temperature-like
quantities—to be modulated by the state of the
entire field. As a result, these systems naturally exhibit complex
collective behaviour, including self-organised criticality, emergent
phase transitions, and rich dynamical phenomena not present in purely
local theories. The interplay between local dynamics and global
feedback encapsulates a new paradigm for modelling statistical
systems, with potential applications ranging from condensed matter
physics to biological and social systems.

\medskip  
In the standard Ising model on the line with open boundary conditions, the degrees of freedom are spins $z_i \in \{-1,+1\}, i=1 \dots L$, where $L$ is the total number of spins. The {\it
  local} standard action is given by the 
standard nearest neighbour interaction:
$$
S \; \propto \; \sum _{i=1..L-1} z_i z_{i+1} \; .
$$
In this paper, we will explore versions of Ising models, the actions of which are
(simple) functions of $S$, such as
$$
S^2 \; \propto \left ( \sum _{i=1..L-1} z_i z_{i+1} \right )^2 =\; \sum _{i,j=1..L-1}z_i z_{i+1} z_j
z_{j+1} \; .  
$$ 
This induces a type of non-locality where all
spins interact with each other.
We motivate the study of models with actions of this form below by showing that this type of interaction can occur naturally in many physical systems. 

\medskip
Denote the action of a field theory by 
$S(\theta; \kappa )$, where
$\theta (x)$ encapsulates the degrees of freedom in 1-dimension
spanned by the variable $x$, and $\kappa$ encapsulates the model parameters. If $S$ is chosen so that
$S(\theta; \kappa )$, has only short range interactions for the field
$\theta $ for given parameters $\kappa $, non-locality can be induced
by introducing a dependence of the parameters $\kappa $ on observables
(expectation values) of the field theory such as action density or
magnetisation. Examples include models for social dynamics or opinion
models, where individuals modify their openness to influence (noise
level) based on the perceived consensus~\cite{Sirbu2017}. 

\medskip  
To illustrate how non-locality arises, we will explore in more detail the example of feedback-driven effective temperatures in protein
thermodynamics. We adopt a rigid polymer model where polymers are
described by a set of angles $\theta _1 \ldots \theta _L$ specifying
the position of rigid links along the polymer chain (see
figure~\ref{fig:0}). A typical local action is given by
$$
S \; = \; \sum _{a, i=1..L-1} \beta \; \cos \Bigl( \theta
^{(a)}_{i+1} -  \theta ^{(a)}_i \Bigr) \; + \; V(\theta ) , 
$$
where $i$ labels the position and $a$ the polymer in a multi-polymer
ensemble. The parameter $\beta $ describes stiffness. For low $\beta
$, the angles are (almost randomly) oriented leading to a low action
``crumpled-up'' state of the polymers, whereas for large  $\beta $,
the state with $\theta _i = \, $constant is preferred leading to
``stretched-out'' polymer ensembles. The potential $V$ ensures that
the polymers are confined to the encasing (shown as the boxes in
figure~\ref{fig:0} (middle)). More interaction terms could be added that ensure
self-avoidance. 

\medskip  
Let us now introduce mesoscopic feedback by
considering polymers in a dense environment. If the polymers are in a low
action state and if each of them is mainly confined to a sphere then, with
self-avoidance and densely packed spheres, it is more difficult for the
polymers to move to the higher action state of a linear
configuration. Consequently, the effective stiffness parameter should be
lower. Conversely, if the polymers are in a  high action, linear state, it
would be more difficult in a dense environment to reach a crumpled
state and consequently the effective stiffness parameter should be higher. 

\medskip  
Rather than working with a microscopic theory of many polymers,  if we are only
interested in their geometry we can
switch to an effective description of a 1-polymer state where the stiffness parameter is now taken as an increasing function of the {\it action density}
$$
s \; := \; \frac{1}{L-1} \sum _{ i=1..L-1} \; \cos \Bigl( \theta 
_{i+1} -  \theta _i \Bigr)  \;,
$$
for example $\beta (s) = \kappa s$. Note that this is an intensive quantity as needed for the stiffness. 
A simple effective model for the angles
$\theta _1 \ldots \theta _L$ of a single polymer in a dense environment is then given by 
\bea 
S_\mathrm{eff} &=& \sum _{i=1..L-1} \beta (s) \; \cos \Bigl( \theta
_{i+1} -  \theta _i \Bigr) , 
\nonumber \\
 &=&  \frac{\kappa}{L-1} \left ( \sum _{ i=1..L-1} \; \cos \Bigl( \theta 
_{i+1} -  \theta _i \Bigr)  \right )^2\; . 
\nonumber
\ena

\medskip  
We next discuss a second motivation for a non-local 1D field
theory, which belongs to the large class of models that emerge from
higher dimensional field theory. These emerging theories (QFTs) are
central to understanding holography, dimensional reduction, and
effective field theories. Examples are rooted in both
string theory~\cite{Maldacena1999,witten1998antisitterspaceholography} 
and condensed matter physics~\cite{PhysRevLett.101.031601}, where
low-dimensional QFTs emerge through compactification, boundary
effects, or dualities.

\medskip
To illustrate how a mesoscopic type of interaction with infinite range
emerges from a local 3D field theory, we consider Ising spins $z_i = \pm
1$, $i=1 \ldots 3$ that exist in the 1-dimensional subdomain at the
the surface of a 
3-dimensional cube (see figure~\ref{fig:0}, right panel). The spins
interact with the field $J_\ell $ associated with the links of the
cube. For illustration purposes only, our theory is described by the action:
\bea 
S & =& \sum _{\ell \in (x,y) } z_x J_\ell z_y \; - \; \frac{1}{4\kappa }
\sum _{\ell}
J_\ell ^2 \; - \; \kappa _2 \, \sum _{x, (\ell_1 \ell_2 \ell_3 ) \in x}  
\nonumber \\
&& \left[ \frac{ ( J_{\ell_1 } -
  J_{\ell_2 })^2 }{ J_{\ell_1 }^2 +  J_{\ell_2 }^2 } + \frac{ ( J_{\ell_1 } -
  J_{\ell_3 })^2 }{ J_{\ell_1 }^2 +  J_{\ell_3 }^2 } + \frac{ ( J_{\ell_2 } -
  J_{\ell_3 })^2 }{ J_{\ell_2 }^2 +  J_{\ell_3 }^2 } \right] \; . 
\nonumber 
\ena 
The first sum extends over all links $\ell $  of the lattice, and $x$ and  $y$ are the lattice sites that are located
at the boundary of the elementary link $\ell $, the second sum extends
over all links $\ell $, and  in
the third sum $\ell_1 \ldots \ell_3$ are the three elementary links that ``belong'' to a
particular site $x$. The theory has only local interactions. It is
described by the partition function:
$$
Z \; = \; \sum _{\{z\}} \int _J \; \exp \Bigl\{ S \Bigr\} \; . 
$$
The 1D effective theory emerges by ``integrating out'' the interaction
field $J_\ell $. For the purpose of this illustration we are interested in two limits: firstly, for
$\kappa _2 =0$, the integral is Gaussian and can be performed by
completing the square. By virtue of $(z_xz_y)^2 =1 $, we observe that
the emerging action for the spin is constant, i.e., the effective
spin theory is that of random spins. Secondly, in the limit $\kappa _2
\to \infty $, we find that fluctuations of the interaction field $J$
are suppressed though not its scale, i.e.,
$$
J_{1} = J_{2} = J_{3} = \ldots =: J \; . 
$$
We find in this limit:
$$
S \; = \; J \; \sum _{x,y \in \ell } z_x z_y \; - \;
\frac{L-1}{4\kappa } \, J^2 \; . 
$$
Integrating over the scale $J$, we find that the emerging
1-dimensional field theory has infinite range but is of the mild
mesoscopic type:
$$
S _\mathrm{eff} \; = \; \frac{ \kappa }{ L-1 } \left( \sum _{x,y, \in
    l} z_x z_y \right)^2 \; . 
$$

\bigskip
\section{Phase transitions in Ising like models in 1D \label{sec:ising}}
\label{sec:phase}

\subsection{Standard Ising Model} 

\begin{figure*}
\includegraphics[height=5cm]{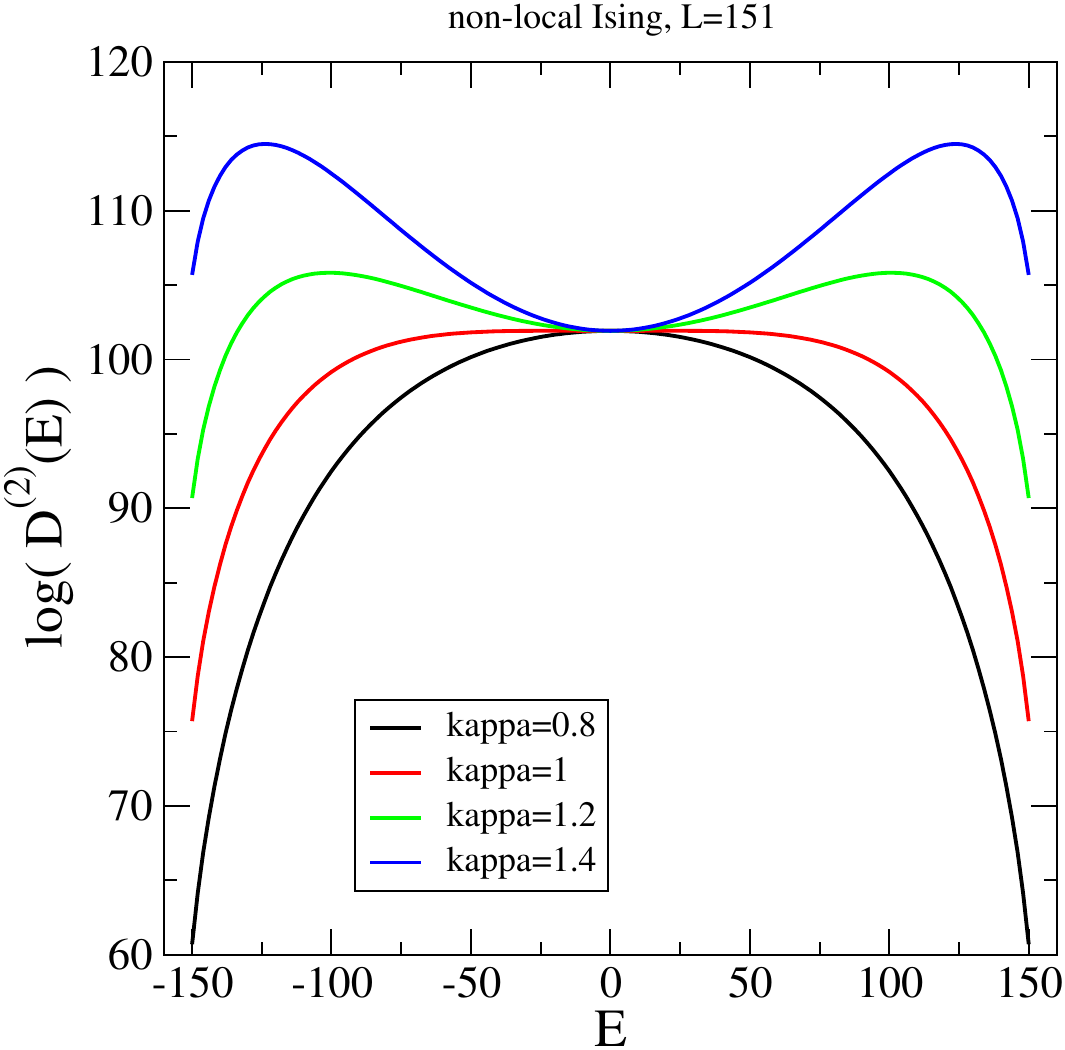}  \hspace{0.5cm}
\includegraphics[height=5cm]{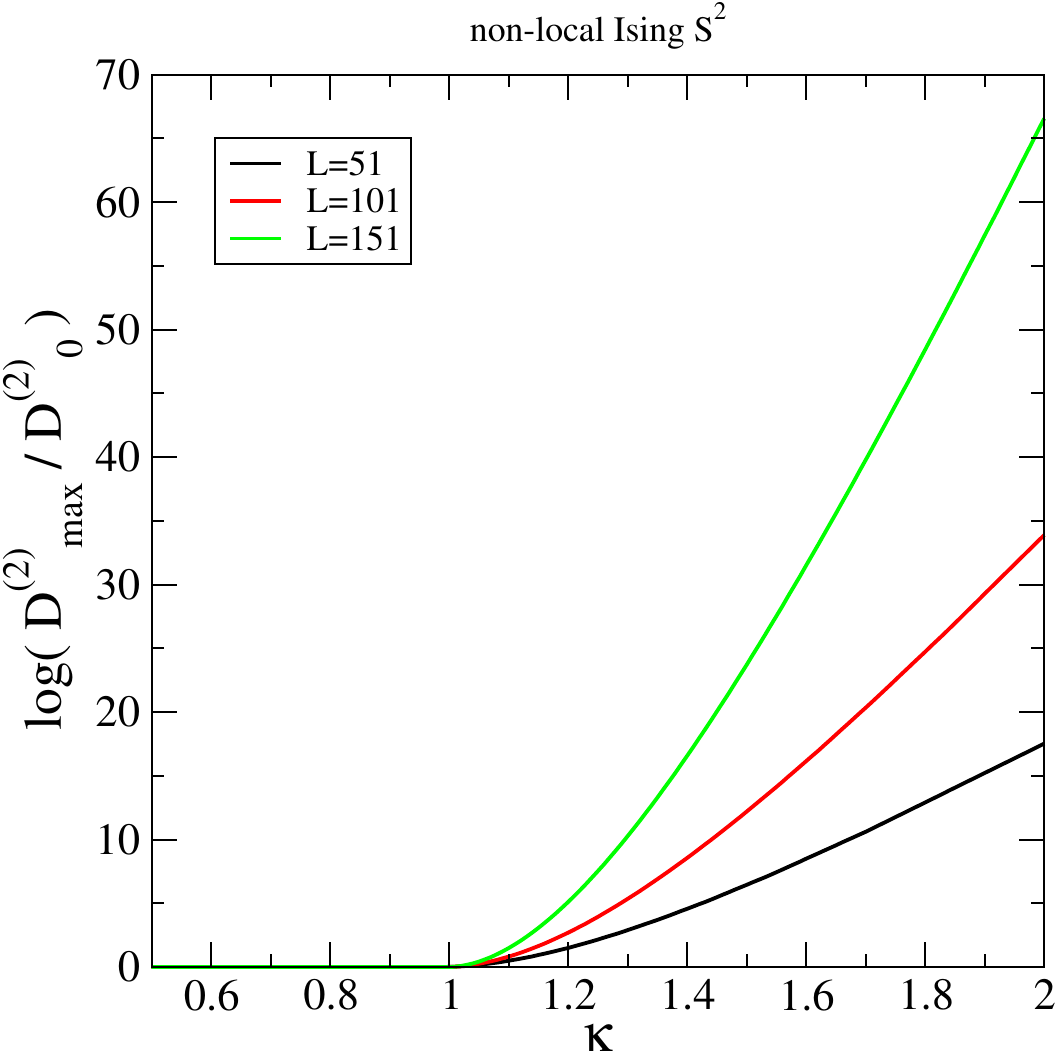}\hspace{0.5cm}
\includegraphics[height=5cm]{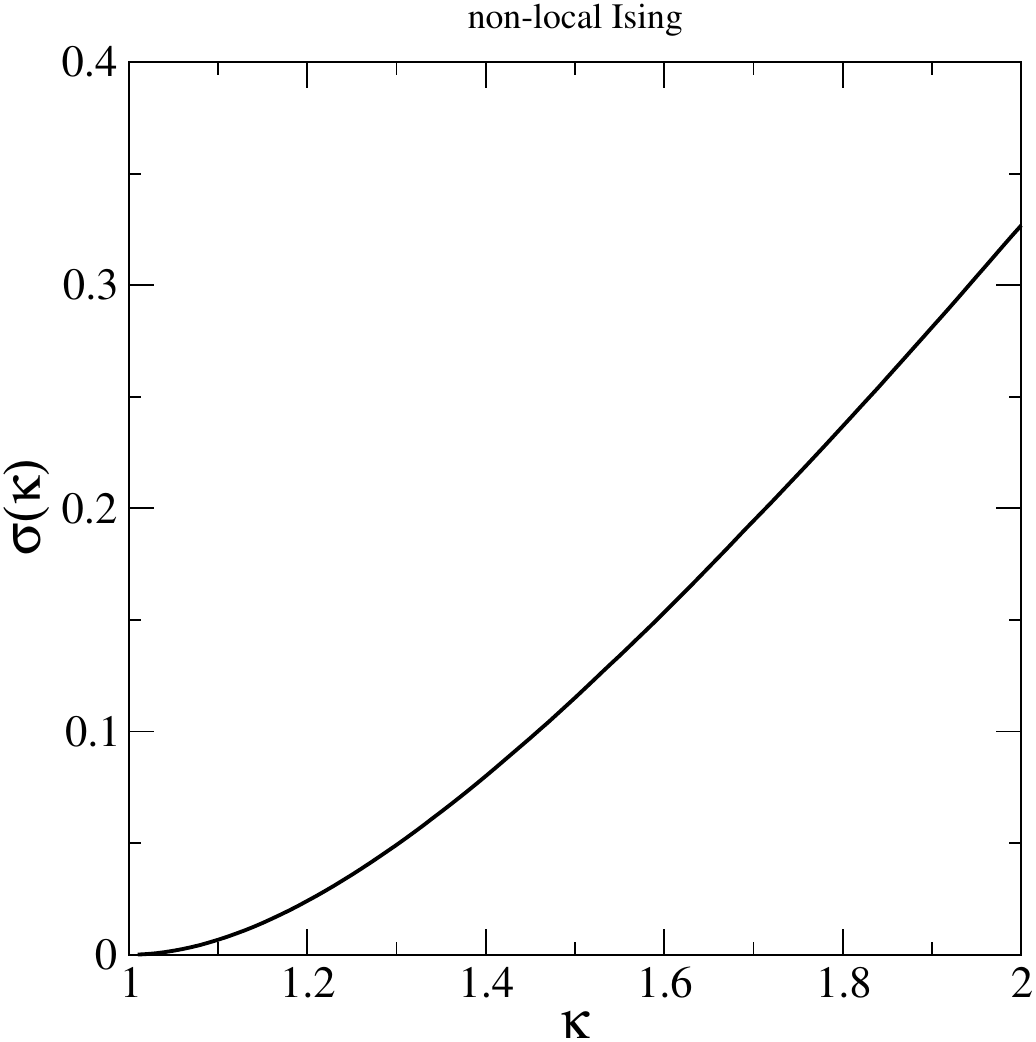} 

  \caption{\label{fig:1} Left: Marginal distribution for the
    nearest-neighbour interaction $E$
    (non-local interaction) for several interactions strengths $\kappa
    $. Middle: Maximum marginal probability over that at $E=0$ for
    several extensions $L$.  Right: Interface tension $\sigma $ as a
    function of the coupling 
    strengths $\kappa $. }
\end{figure*}
Recall the standard nearest-neighbour Ising model with  action given by:
\be
S \; = \; \sum _{l \in \langle xy\rangle } z_x z_y \; ,
\label{eq:2.1}
\en
where $\ell $ is an elementary link joining neighbouring sites $x$ and
$y$. We start the exploration with a calculation of the
density of states for the interaction $S$ from \eqref{eq:2.1}: 
\be
\rho (E) \; = \; \sum _{\{z\}} \delta \Bigl( E, S[z] \Bigr) \; ,
\label{eq:2.2}
\en
where $\delta $ is the Kronecker symbol. This can be evaluated in closed
form by a shift of variables:
$$
z_x \, , \; x=1 \ldots L \; \; \;  \Rightarrow \; \; \;  z_1, \; u_x , \; x=1
\ldots L-1
$$
where
$$
u_x \; = \; \left\{
  \begin{array}{cc} 1 & \hbox{for} \; \; z_x = z_{x+1} \\
    -1 & \hbox{for} \; \; z_x \not= z_{x+1}
  \end{array} \right. \; . 
$$
The action only depends on $u$:
$$
S \; = \; \sum _{x=1}^{L-1} u_x \;  = \; n - (L-1-n) \; = 2n - L + 1
\; , 
$$
where $n \in [0,L-1]$ is the number $u$-variables that are positive, leaving $L-1-n$
$u$-variables negative. The number of possibilities to render $n$ of our $L-1$
$u$-variables positive is:
$$
\left( \begin{array}{c} L-1 \\ n \end{array} \right) \; . 
$$
The density of states is given by
\be
\rho(E) \; = \; 2 \left( \begin{array}{c} L-1 \\ n \end{array} \right)
\; , \hbo E \; = \; 2n - L + 1 \; .
\label{eq:2.3}
\en
Note that the factor $2$ rises from summing over $z_1$. 

\medskip
With $\rho
(E)$ at our fingertips, we can derive the exact partition function. The same approach also works for models with non-local interactions, which we demonstrate in the next section. 
Inserting the
``$1$'',
$$
\sum _E \delta \Bigl( E, \, S[z] \Bigr) \; = \; 1 \; , \hbo \forall \;
\; \; \{z\}
$$
into the partition function of the standard Ising model, i.e., 
\bea
Z(\beta ) &=& \sum_{\{z\}} \; \exp \left\{ \beta \, S[z]
\right\}
\label{eq:2.6a}  
\ena
we change the order of summations and
substitute $E$ for $S[z]$ by virtue of the $\delta $-function, i.e., 
\bea
Z(\beta ) &=& \sum_E \sum_{\{z\}} \, \delta\Bigl( E, \, S[z]
\Bigr) \; \exp \left\{ \beta\, S[z]
\right\}
\nonumber \\ 
&=&  \sum_E \; \exp \left\{ \beta \, E
\right\} \;\sum_{\{z\}} \, \delta\Bigl( E, \, S[z]
\Bigr)
\nonumber \\ 
&=&   \sum_E \; \exp \left\{ \beta \, E
\right\} \; \rho(E) \; = \;  \sum_E D^{(1)}(E) \; , 
\nonumber 
\ena
where the density of states for the non-local theory can be entirely
expressed by the density $\rho (E)$ for the next-to-nearest neighbours
interaction:
\be
D^{(1)}(E) \; = \; \exp \left\{ \beta \, E
\right\} \; \rho(E) \; .
\label{eq:2.6c}
\en 
As we illustrate in Section \ref{sec:LVL} below, the density $D^{(1)} (E) $ possesses
only a single maximum for all finite values of the inverse temperature
$\beta $. We do not face the situation where two distinct values of $E$
are maxima and equally likely, which signals the presence of a phase
transition. We thus recover the familiar result that phase transitions are absent
in a 1D statistical field theory with local interactions.

\subsection{Ising model with $S^2$ non-locality \label{sec:IsingS2}} 

\begin{figure*}
\includegraphics[height=5cm]{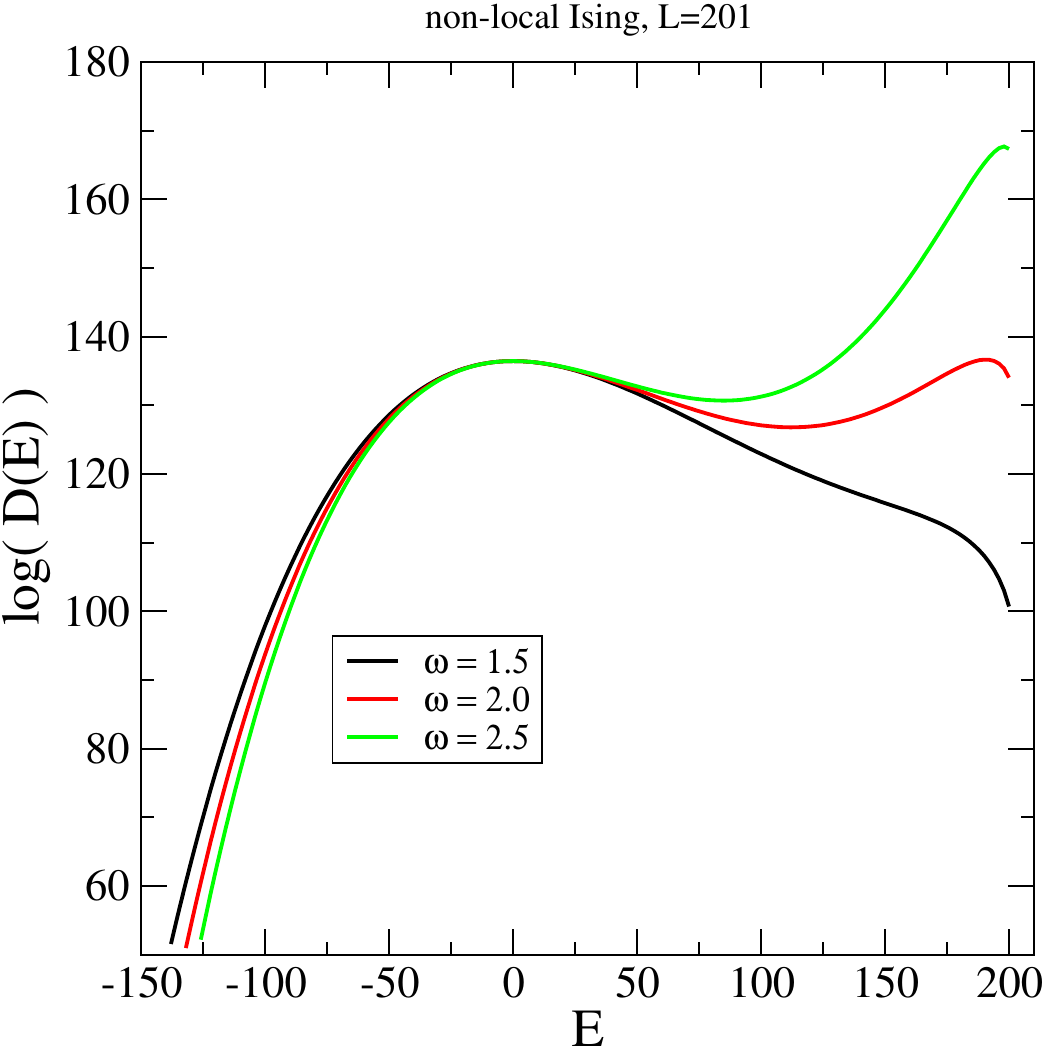}  \hspace{0.5cm}
\includegraphics[height=5cm]{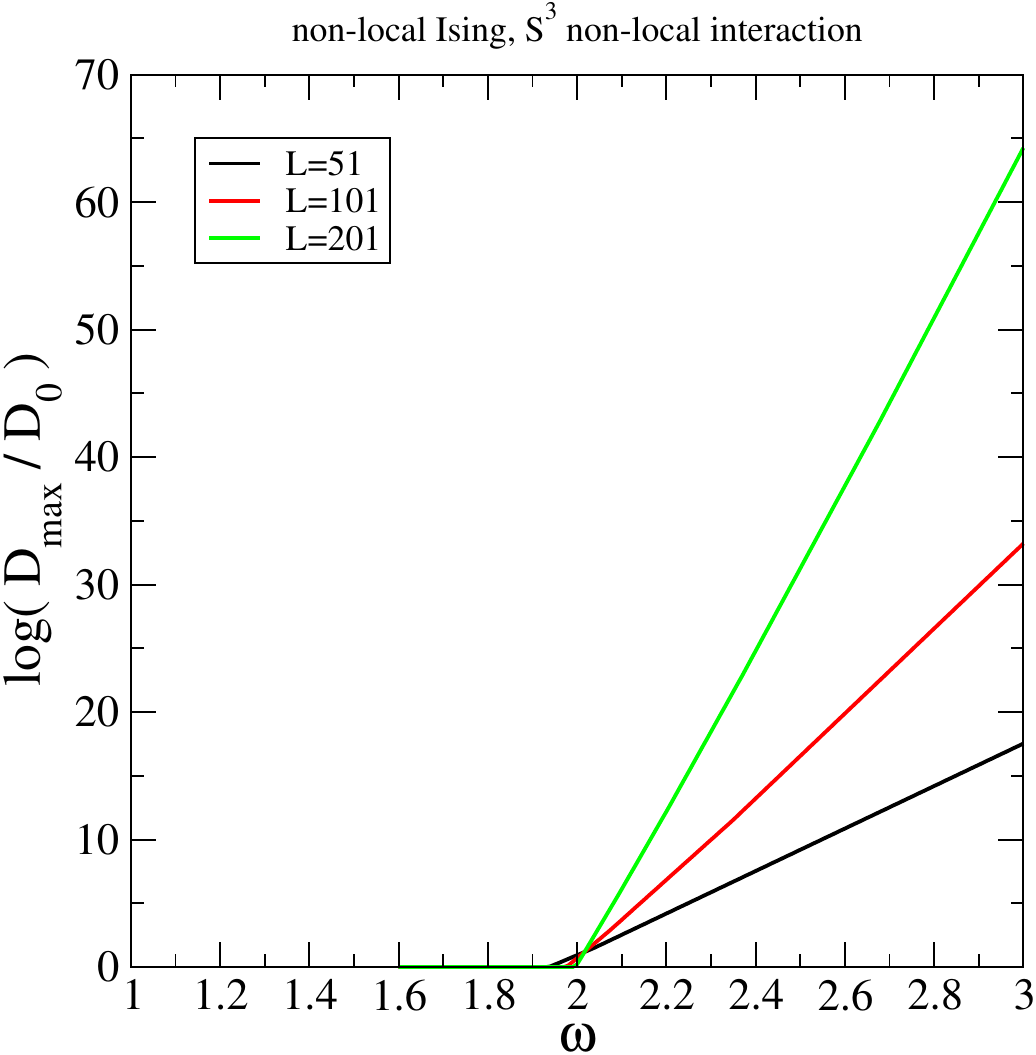}  \hspace{0.5cm}
\includegraphics[height=5cm]{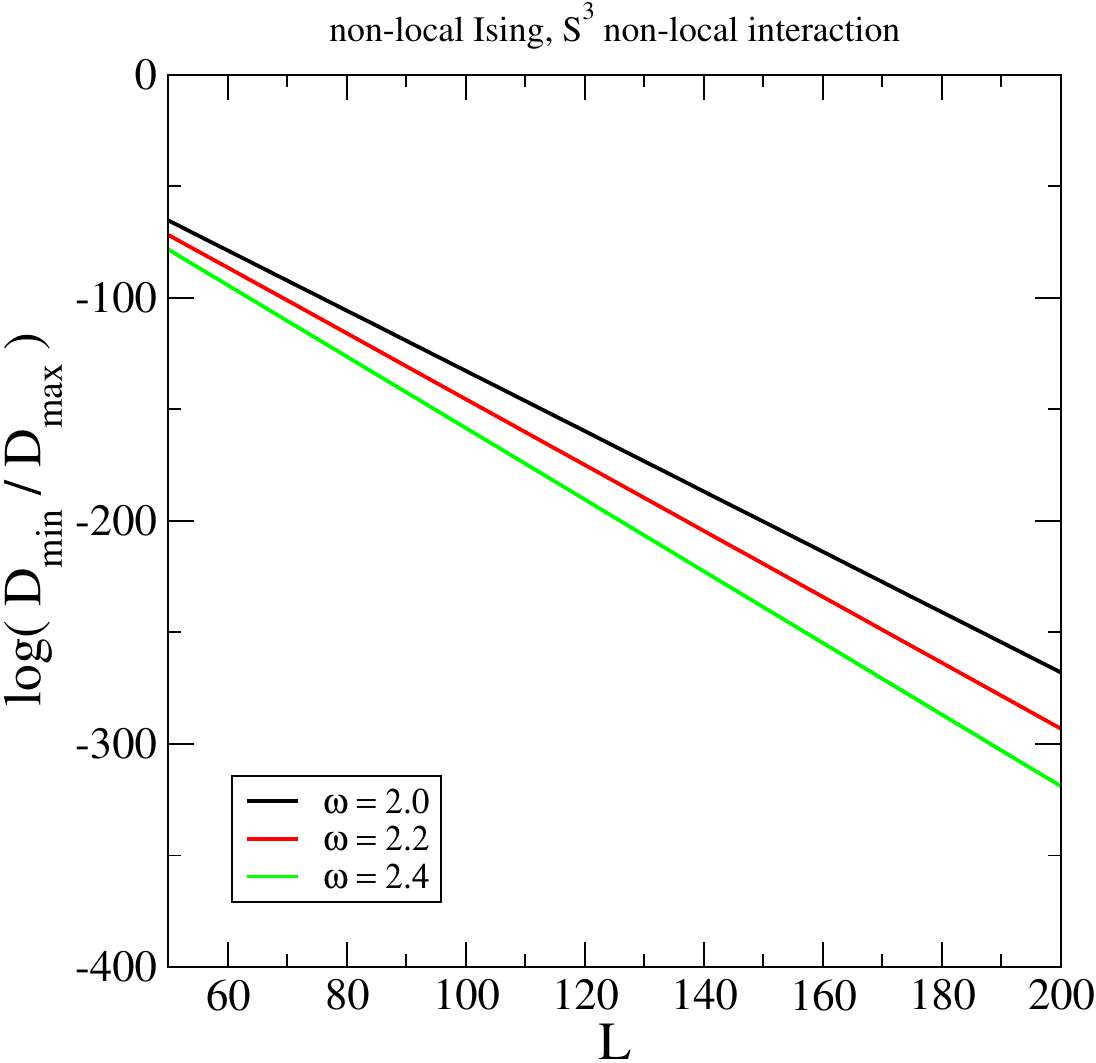} 
  \caption{\label{fig:2} Left: Distribution marginal
    distribution for the nearest neighbour interaction  $E$. Middle:
    Suppression of the symmetric state $E=0$ in the broken phase. 
    Right: Gap formation in the limit $L \to \infty$.  }
\end{figure*}  

We now show how the same approach can be used 
for the model with the $S^2$ action:
\be
S^2 \; = \; \frac{1}{L-1} \sum _{\langle xy\rangle, \langle uv \rangle }z_x z_y z_u
z_v \; .  
\label{eq:2.1b}
\en 
The partition function we consider here is given by
\be
Z(\kappa ) = \sum_{\{z\}} \exp \left\{ \frac{ \kappa }{2 (L-1)} \, S^2[z]
  \right\} \; ,
\label{eq:2.6}
\en
where $S$ is the nearest neighbour interaction in (\ref{eq:2.1}). We can
consider the interaction between those spins depending on a medium and
the interaction strength influenced by the {\it action density }
$S/(L-1)$:
$$
\beta(S) \, S \; , \hbo \hbox{with} \; \; \beta(S) = \frac{ \kappa }{2}  \,
\frac{ S}{L-1} \; . 
$$
If more and more spins align, e.g., in a ferromagnetic setting, the
effective interaction $\beta $ between spins even increases
further. The present model has a ferromagnetic antiferromagnetic
symmetry (apart from the usual spin-flip symmetry of the standard
Ising model). The transformation 
\bea 
z _x \to z _x && \; \; \; \hbox{for} \; \;  x \; \; \hbox{even},
\nonumber \\
z _x \to - z _x && \; \; \;\hbox{for} \; \;  x \;
\;\hbox{odd}, 
\nonumber
\ena 
maps $S \to -S $. This is a symmetry since the probabilistic measure
only depends on $S^2$.

\medskip
Following the steps of the previous subsection, we express the density
of states of the $S^2$ model in terms of the density $\rho (E)$ of the
nearest-neighbour-interaction $E$\footnote{This is {\it
    not} the action of the new $S^2$ model. It, however, still provides
  useful insights in  the systems behaviour.}: 
\be
D^{(2)} (E) \; = \; \rho(E) \; \exp \left\{ \frac{ \kappa }{2 (L-1)} \, E^2
  \right\} \; . 
\label{eq:2.7}
\en
Because of the ferro- antiferromagnetic symmetry, it is is a
symmetric function in $E$.

\medskip 
Using (\ref{eq:2.3}), the distribution
$E$ can be easily calculated numerically. The result is shown in
figure~\ref{fig:1} for several values of $\kappa $. We observe that if
the interaction strength $\kappa $ exceeds a critical value $\kappa
_c \approx 1$, the most likely action is obtained for $\pm E $,
$E\not=0$. This indicates the spontaneous breakdown of the
ferromagnetic antiferromagnetic symmetry. This would happen if the
transition probability between the two possible ground states, i.e.,
ferromagnet and anti-ferromagnet, vanishes in the infinite volume limit
$L \to \infty $. To explore this idea further, we study the marginal
distribution $D^{(2)}(E=0)$ over $D^{(2)}_\mathrm{max}$ for three
values $\kappa > \kappa_c$ as a function of 
$L$. We observe an exponential suppression of the probability:
\be
D^{(2)}(E=0)/D^{(2)}_\mathrm{max} \; = \; \e ^{-\sigma (\kappa ) \, L } \; .
\label{eq:2.8}
\en
This suppression occurs from faults in the (anti-) ferromagnetic
order, and $\sigma $ is called {\it interface
  tension}. If we dare to extrapolate without proof to the limit $L\to
\infty $, we indeed observe the spontaneous breaking of the
ferromagnetic antiferromagnetic symmetry. Figure~\ref{fig:1}, right
panel, shows the behaviour of the interface tension as a function
of $\kappa $. The tension vanishes at the critical coupling $\kappa
_c$ and monotonically increases with $\kappa $ from there on. The transition is 2nd
order. 

\subsection{Ising model with $S^3$ non-locality \label{sec:IsingS3}} 

The $S^3$ model possesses only the spin-flip symmetry, which is
familiar form the standard local Ising model, i.e.,
$$
z _x \to - z _x \; , \hbo \forall x \; . 
$$
While the standard 1D Ising model does not have a spontaneous breaking
of a symmetry, the interesting question is whether a non-local version
that shares the same symmetry pattern can break the symmetry
spontaneously.

\medskip 
\begin{figure*}
\includegraphics[height=5cm]{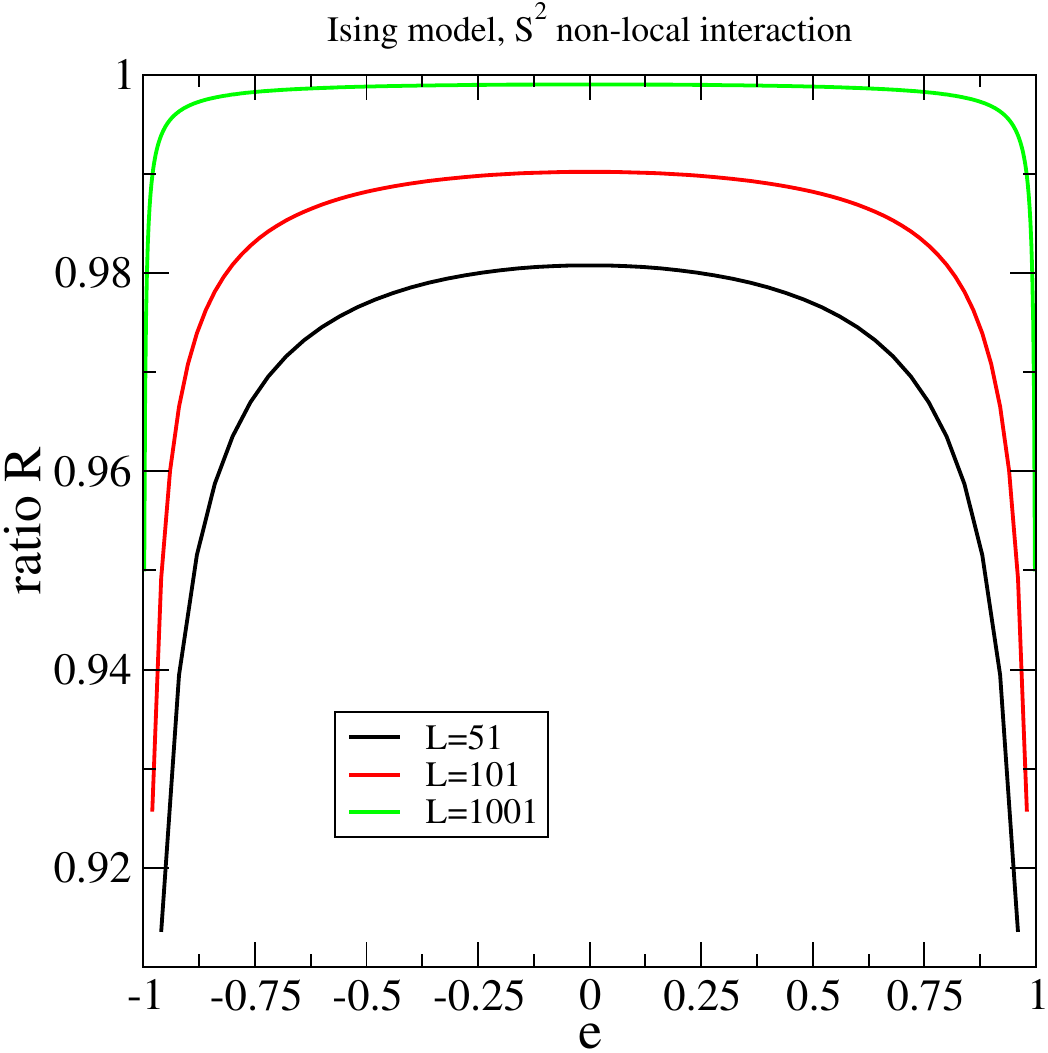}  \hspace{0.5cm}
\includegraphics[height=5cm]{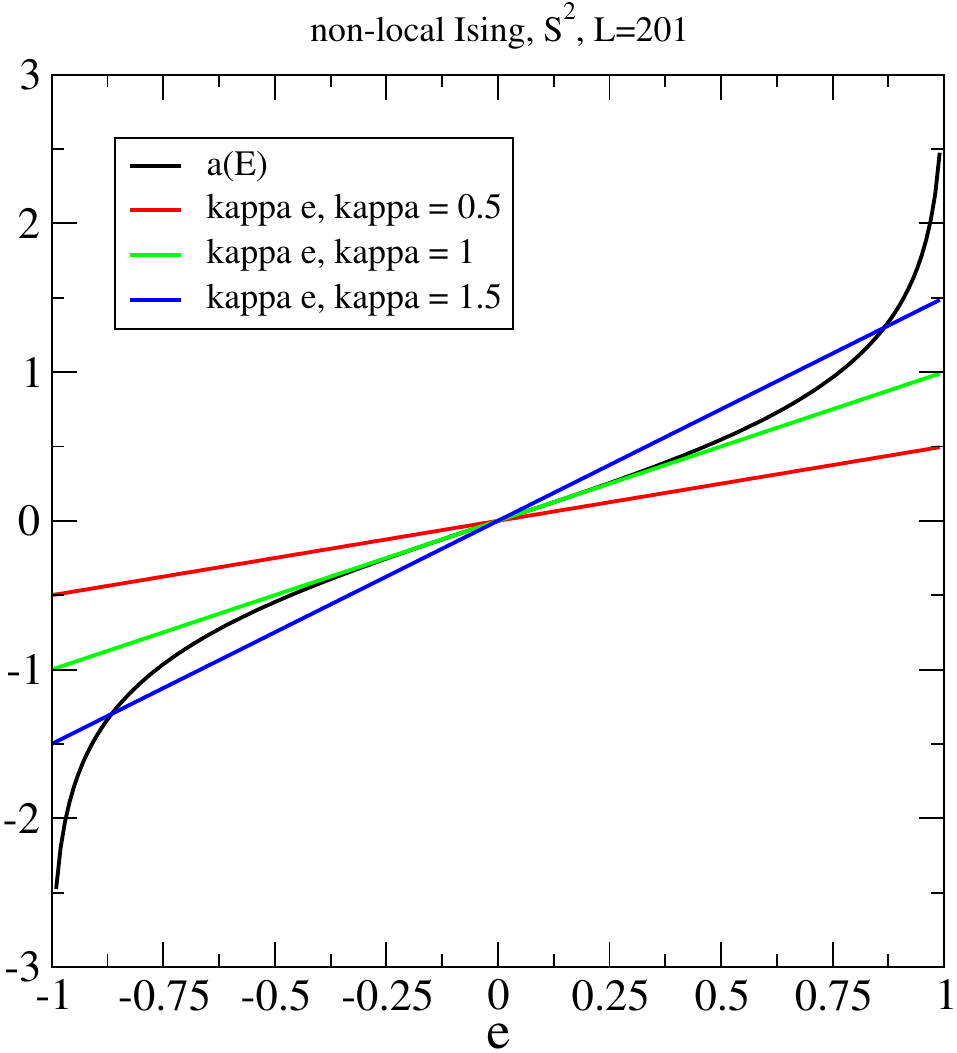}  \hspace{0.5cm}
\includegraphics[height=5cm]{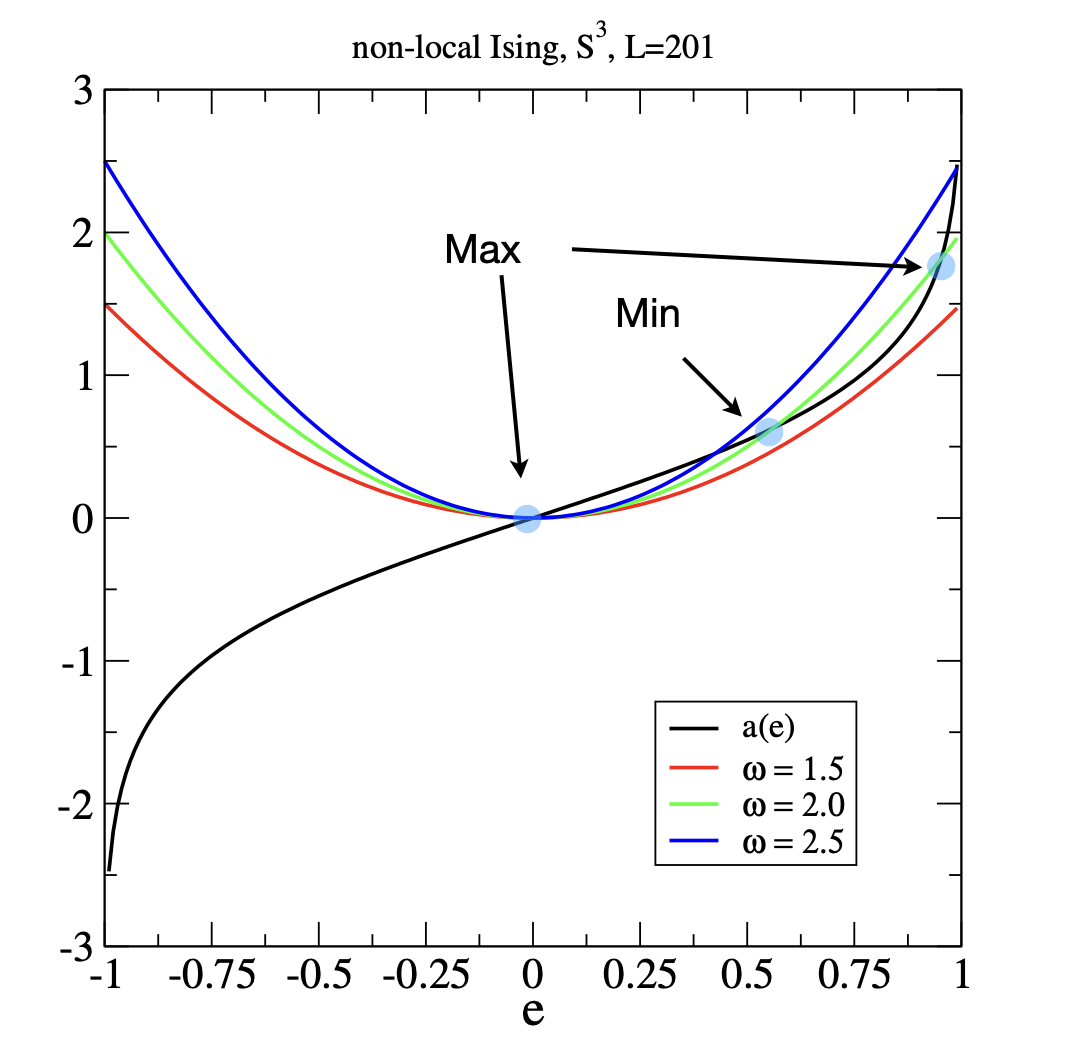} 
  \caption{\label{fig:3} Left: LLR coefficient at finite volume over
    its asymptotic value (\ref{eq:2.14}). Screening for phase
    transitions using the LLR coefficient for the the $S^2$ (middle)
    and $S^3$ (right) case. }
\end{figure*}
The $S^3$ model can be understood as an Ising model with effective
interaction induced by a medium that is sensitive to the energy stored
in the spin system. This time the effective interaction, i.e., 
$$
\beta (S) \; = \; \frac{\omega }{3} \; \left( \frac{ S }{L-1}
\right)^2 \; , 
$$
is
proportional to $S^2$ and the partition function is given by:
\bea
Z(\omega ) &=& \sum_{\{z\}} \exp \left\{ \beta \Bigl(S[z]\Bigr)  \, S[z] \right\}
\nonumber \\ 
&=& \sum_{\{z\}} \exp \left\{ \frac{ \omega }{3 (L-1)^2} \, S^3[z]
  \right\} \; .
\label{eq:3.1}
\ena
It is ideal to use the nearest neighbour interaction $S$ as a
collective coordinate since it is a good litmus test for symmetry
breaking. In the unbroken so-called {\it paramagnetic phase}, the
disorder in a particular spin configuration implies $\langle S \rangle
= 0 $ already upon the spatial average. The variable $S$ is also
insensitive to the spin-flip: 
$$
z \to - z \hbo \rightarrow \hbo S \to S \; . 
$$
In the {\it ferromagnetic phase} and even for the finite volume
ergodic stochastic simulation, we expect $\langle \not= 0 \rangle $.
The partition function again can be expressed in closed form as an
integral over the the nearest-neighbour
interaction $E$:
\bea
Z (\omega ) &=& \sum _E D^{(3)} (E) \; ,
\nonumber \\
D^{(3)} (E) &=& \rho(E) \; \exp \left\{ \frac{ \omega }{3 (L-1)^2} \, E^3
  \right\} \; , 
\label{eq:3.2}
\ena
with $\rho (E)$ already calculated in (\ref{eq:2.3}).
For small enough $\omega $,  the distribution $D^{(3)} (E)$ has the
global maximum at $E=0$ (see figure~\ref{fig:2}). Even at finite
volumes, the most likely state is ${\it paramagnetic}$. We find that for
sufficiently large $\kappa $, the distribution $D^{(3)} (E)$  develops
a second local maximum at large $E>0$. The local maximum becomes the
global one for $\kappa > \kappa _c \approx 2.0$. The value for
$ \kappa _c $ slightly depends on the volume $L$ but seems to converge
to a finite value with increasing $L$. If we consider the rift between
the maximum of the marginal distribution $D^{(3)}(E)$ and its minimum, it
increases exponentially with increasing $L$  (see figure~\ref{fig:2},
right panel), suggesting that the reflection symmetry is indeed
spontaneously broken in the infinite volume limit. Here, we observe
that the transition is 1st order. 

\subsection{The large volume limit  }
\label{sec:LVL}

Let us further explore the phase transitions of the previous
subsections in particular the behaviour of order under an increase of
the volume $L$. To this aim, we introduce the LLR coefficient $a(E)$
by~\cite{Langfeld:2012ah,Langfeld:2015fua}
\be
a (E) \; = \; - \; \frac{d}{dE} \, \ln \rho(E) \; .
\label{eq:2.10}
\en
(We use the notation $a(E; L)$ for this quantity where it is useful to emphasise the dependence on $L$.)
This derived quantity is {\it intensive}, i.e., there exists a well-defined limiting function $\bar{a}: ]-1,1[ / \{0\} \to \mathbb{R}$ satisfying 
$$
\lim_{L, E \to \infty}a(E; L) =  \bar{a} ( e ), 
$$
where the limit is taken under the restriction $E/(L-1) \to e$. 

Let us now consider the distribution $D^{(2)}(E)$ for the theory with
the $S^2$ action, which can be written in terms of the LLR coefficient
as 
\be
D^{(2)}(E) \; = \; D^{(2)}(0) \; \exp \left\{ - \int _0^E \left[ a(E^\prime) -
    \frac{\kappa }{L-1} \, E^\prime  \right] \; dE^\prime \right\} \; .
\label{eq:2.11}
\en
Changing the variable from the extensive action $E$ to the interaction
density $e := E/(L-1)$, we find:
\be
D^{(2)}(E) \; = \; D^{(2)}(0) \; \exp \left\{ - (L-1) \int _0^e \Bigl[ \bar{a}(e^\prime) -
  \kappa  \, e^\prime  \Bigr] \; de^\prime \right\} \; . 
\label{eq:2.12}
\en
We recover the familiar result from thermodynamics that probability densities
for aggregate variables sharply peak around their most likely
values. Extrema are obtained by 
\be
\frac{dD^{(2)}(E)}{dE} =0 \; \; \; \Rightarrow \; \;
\bar{a}(e) \; = \;    \kappa  \, e \; . 
\label{eq:2.12b}
\en
For the Ising model (\ref{eq:2.3}), the energies are discrete and we define the LLR
coefficient by a centred difference:
\bea
a(E) &=& -\frac{1}{4} \; \left[ \ln \, \left( \begin{array}{c} L-1
\\ n+1 \end{array} \right) \; - \; \ln \, \left( \begin{array}{c} L-1
\\ n-1 \end{array} \right) \right]
\nonumber \\
&=& - \frac{1}{4} \,  \ln \, \left( \frac{L}{n+1} - 1 \right) \;
- \; \frac{1}{4} \,  \ln \, \left( \frac{L}{n} - 1 \right) \;
\label{eq:2.13}  \\
E &=& 2n -L +1 , \hbo n=1 \ldots L-2 \; .
\nonumber 
\ena
Asymptotically for large $L$, we find for the limiting function:
\be
\bar{a}(e) \; =  \frac{1}{2} \; \ln \left( \frac{ 1+e}{1-e}
\right) \; .
\label{eq:2.14}
\en
To explore this limit numerically, we define the ratio
\be
R \; = \frac{a(E;L)}{\bar{a}(e) } \; , \hbo e \in ]-1,1[ / \{0\} \; .
\label{eq:2.5}
\en
Even for moderate sizes such as $L=101$, we find an agreement within
4\% for most of the $e$-domain (see figure~\ref{fig:3}). Deviations
are most significant at the  
boundaries of the domain of support.Let us now use these findings to
screen for phase transitions in the infinite volume limit using (\ref{eq:2.12b}). 

\medskip  
\textbf{Standard local Ising model:} Here, we need to solve
$\bar{a}(e)=\beta $. The function $\bar{a}(e)$ is monotonic leaving us
with only one intersection point for each $\beta $-value. This
provides the expectation value $\langle S \rangle $, which smoothly
moves from negative values ($\beta < 0$) and anti-ferromagnetic
behaviour to ferromagnetic behaviour for $\beta >0$. As in line with
the Mermin-Wagner theorem, we do not encounter a phase transition.

\medskip
\textbf{Non-local Ising model with $S^2$ action:} The crucial
equation here becomes $\bar{a}(e) = \kappa e $. Possible intersection
points are illustrated in figure~$\ref{fig:3}$ (middle). If the
slope $\kappa $ of the straight line is smaller than $1$, the only
intersection point is at $e=0$ and the spin system is in the
paramagnetic phase. For $\kappa > 1$, we find three intersection
points. Those correspond to two maxima and one minimum in the
probability distribution $D^{(2)}$. We observe two distinct maxima: a
ferromagnetic state and its counter part, an anti-ferromagnetic state,
have equal probability (symmetry) and are more likely than the
paramagnetic state at $e=0$. If we start increasing $\kappa $ from low
values, the transition from the one to the two maximum scenario occurs
smoothly at $\kappa = \kappa _c=1$. The transition is second order. 

\medskip
\textbf{Non-local Ising model with $S^3$ action:} We now encounter
$\bar{a}(e) = \omega e^2 $. There is always an intersection point at
the paramagnetic point $e=0$. The question is whether there are
others, and then whether those give rise to a more likely
state. Again, it is easily observed that if $\omega $ is small,
the only maximum is at the paramagnetic point. If $\omega $ exceeds a
critical value, more candidates for the ground state emerge (see
figure~\ref{fig:3}, right picture): while $e=0$ will always be a
maximum, the corresponding value for $D^{(3)}$ at the second maximum
at large $e$ (ferromagnetic state) will decide whether this state is
more likely. The numerical experiments from the previous subsection
indicate that this is indeed the case for $\omega >2$. The transition
is abrupt and 1st order: for $\omega $ slightly below the critical
value, the spin system is paramagnetic $e=0$. With $\omega $ slightly
above $\omega _c$, the system can be already strongly ferromagnetic $e
\approx 1$. 

\section{Criticality of the non-local $S^2$ Ising model}
\label{sec:crit}
\begin{figure}   
\includegraphics[height=5cm]{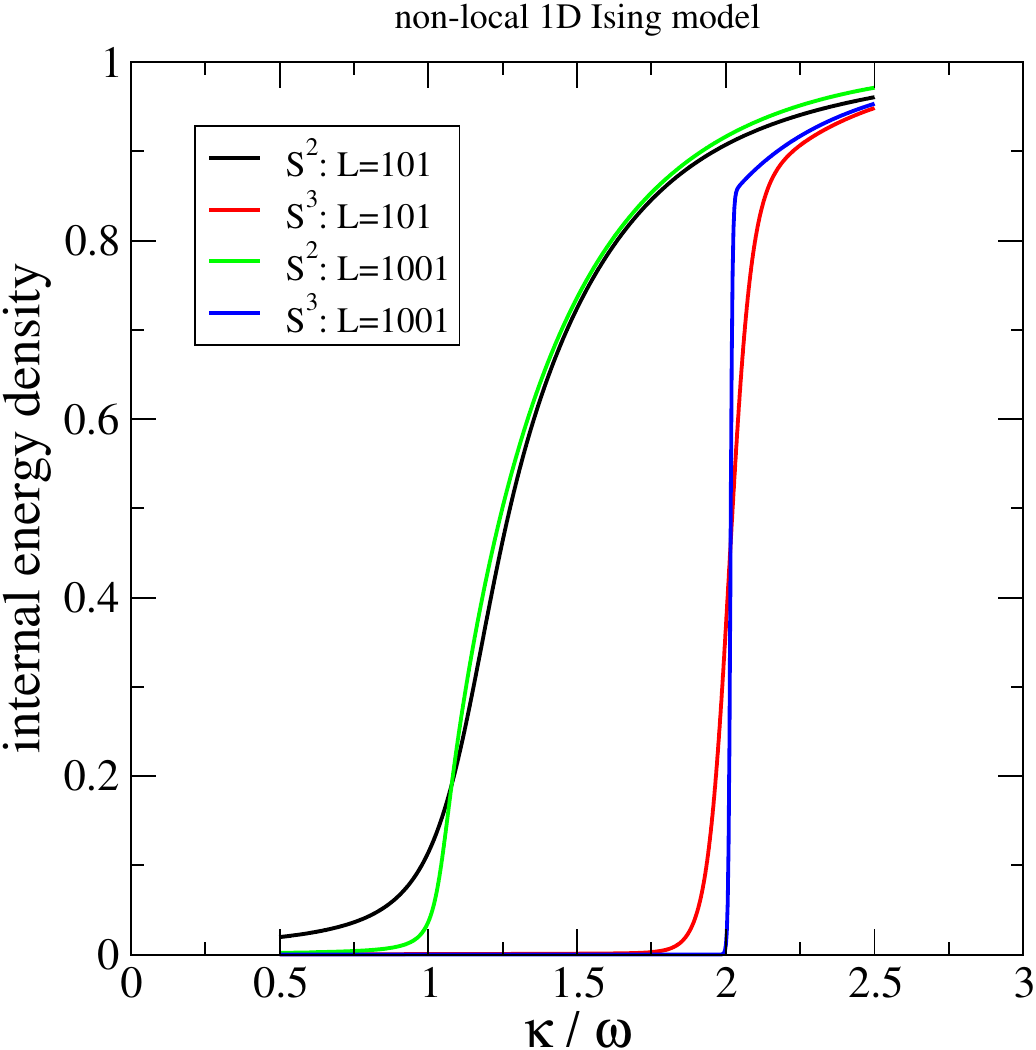}  \hspace{0.5cm}
\caption{\label{fig:4} The internal energy density as a
  function of the couplings for the $S^2$ and $S^3$ theory. 
  }  
\end{figure}
\begin{figure*}  
\includegraphics[height=5cm]{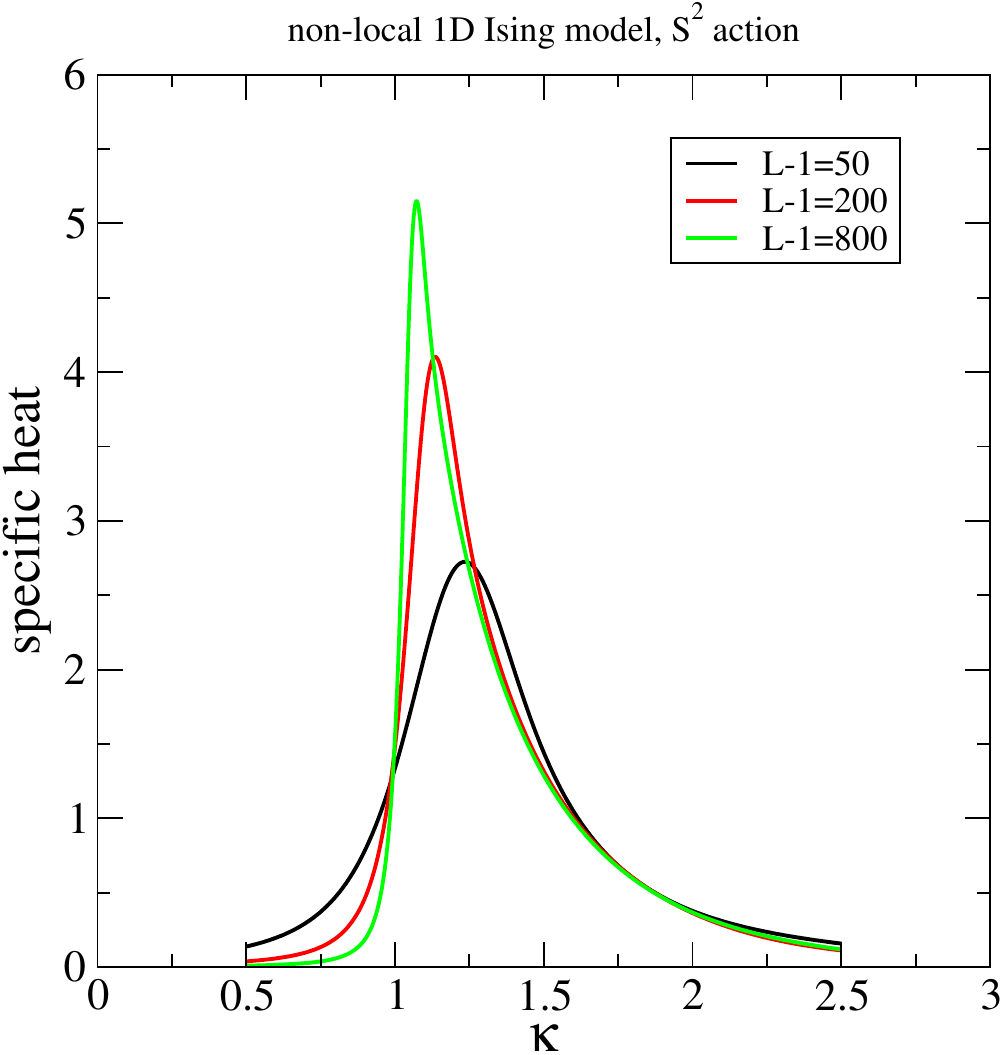}  \hspace{0.5cm}
\includegraphics[height=5cm]{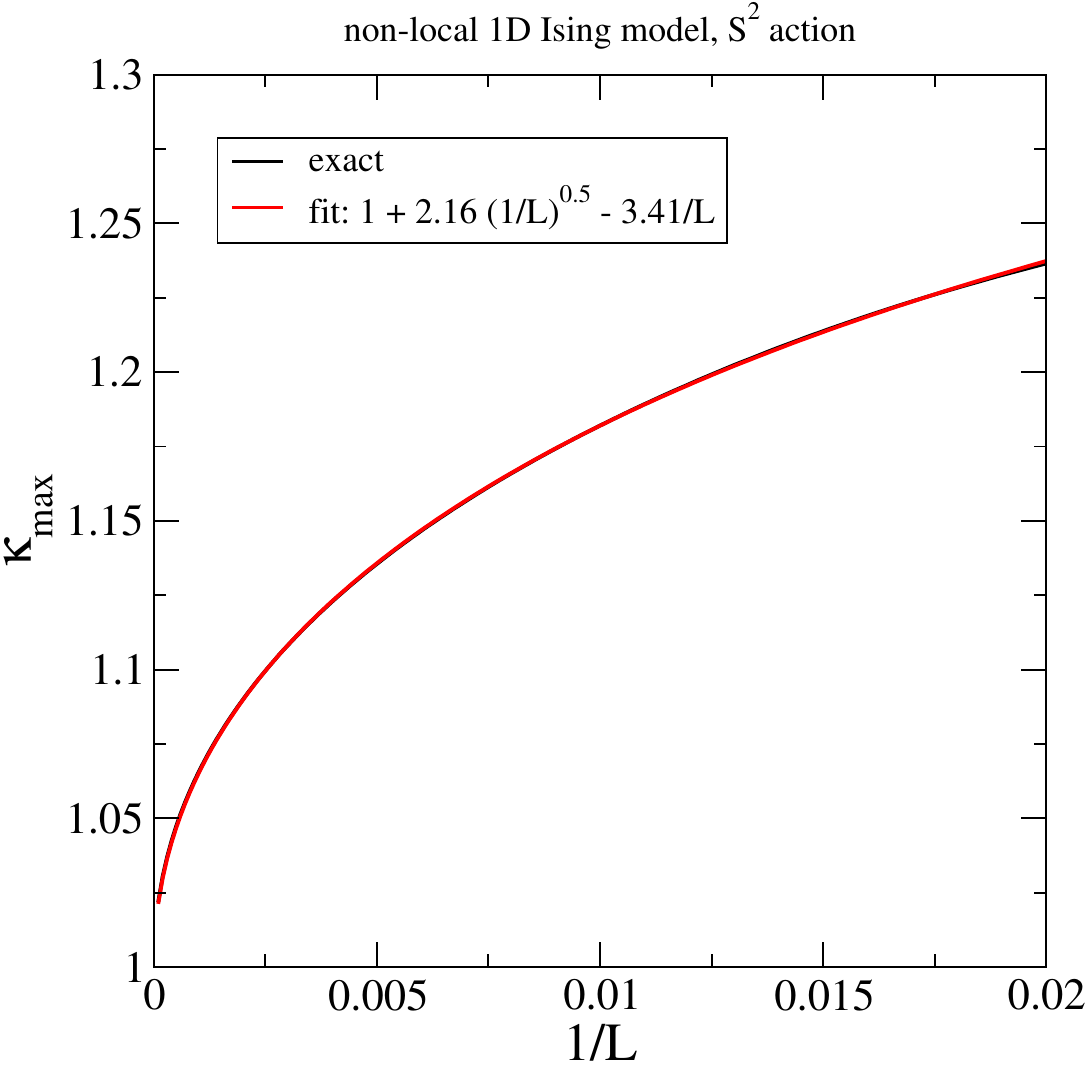}  
\includegraphics[height=5cm]{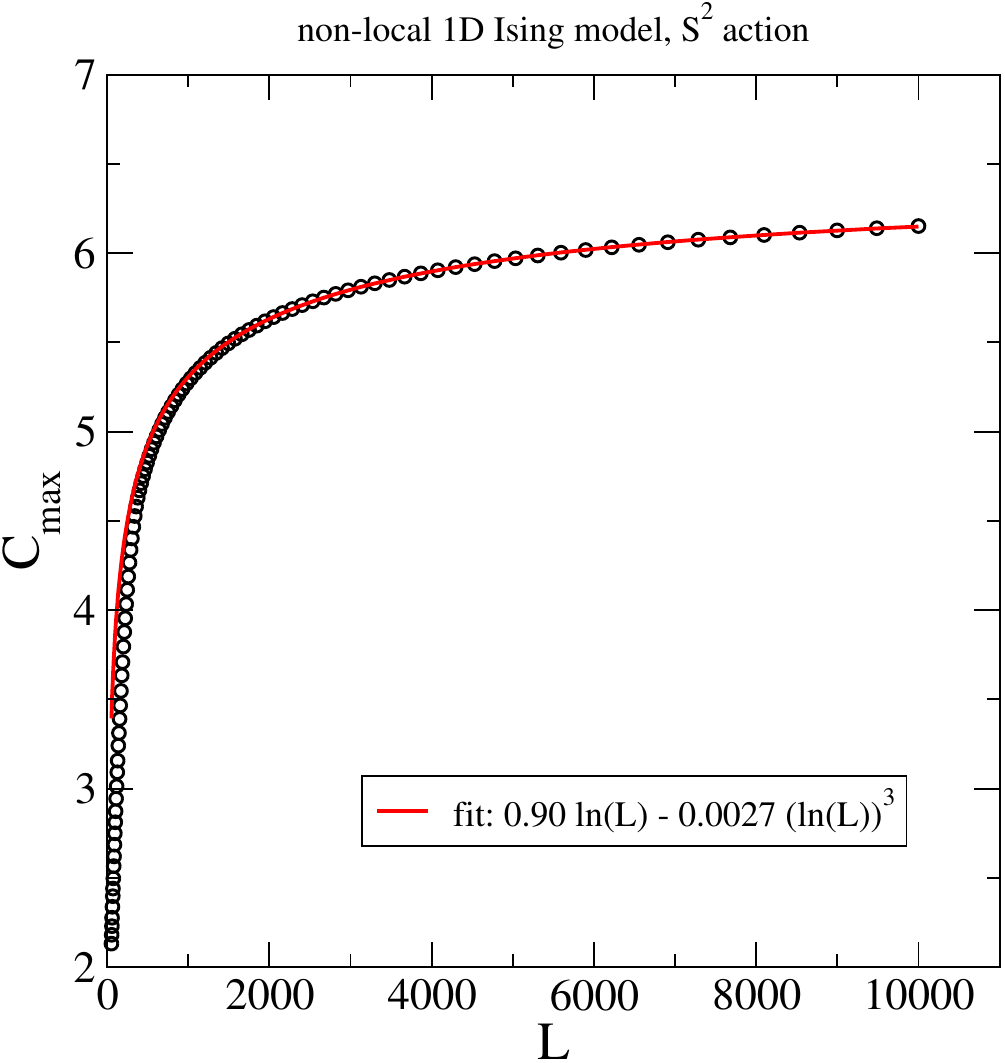}  
  \caption{\label{fig:5} Left: The
    specific heat as function of the coupling strength $\kappa $ for
    several volumes $L$. Middle: Position of the maximum of the
    specific heat as a function of $1/L$.  Right: Scaling of the maximum of the specific
  heat with the system size $L$.  }
\end{figure*}

We now focus on the non-local 1D  Ising model with partition functions
(\ref{eq:2.6}) and (\ref{eq:3.1}) with respective actions
\bea
A_2 &=& \left( \frac{ S }{L-1} \right) \; S  \; ,
\label{eq:4.1} \\
A_3 &=&  \left( \frac{ S }{L-1} \right) ^2 \; S  \; . 
\label{eq:4.2}
\ena 
As explained in the previous subsections, both theories have different
symmetry patterns. In both cases, a spontaneous breakdown of symmetry
occurs, which is 2nd order in the former and 1st order in the latter
case.

\subsection{Internal energy and specific heat}

The internal energy density is defined by
\be
\langle A_\ell \rangle / (L-1) \; , \hbo \ell =2,3 \; .
\label{eq:4.3}
\en
For the $S^2$ theory, we observe typical 2nd order behaviour: the
internal energy density is small in the symmetric phase and then rises
smoothly for couplings $\kappa $ greater than the critical value. For
large volumes $L$, this behaviour is more pronounced. The energy
density develops a cusp in the infinite volume limit but remains
smooth (see figure~\ref{fig:4}). By contrast, the $S^3$ theory shows a sharp rise at the
critical coupling $\omega _c$, the steepness of which increases with
increasing volume. Increasing $L$ from a moderate $101$ to $1001$ lets
us anticipate the 1st order jump of the energy density in the infinite
volume limit (see figure~\ref{fig:4}).

\medskip
In the following, we will focus on the interacting case of the second
order transition of the non-local 1D Ising model with $S^2$ action. 

\medskip
The specific heat $C$ is a further thermodynamical observable with high
phenomenological relevance: 
\be
C_L(\kappa ) \; = \; \left(  \langle A^2 \rangle \; - \; \langle A
  \rangle ^2 \right) \; / \; (L-1) \; .
\label{eq:4.4}
\en
It can be viewed as an integrated correlation function and is therefore
proportional to a power of the correlation length  $\xi $. 
Close to the second order transition, the correlation length $\xi $
diverges with a scaling law of the type
\be
\xi \propto \left\vert 1 - \frac{\kappa }{\kappa _c} \right\vert
  ^{-\nu } \; , \hbo \nu >0 \; ,
\label{eq:4.10}
\en
where $\nu $ is the so-called critical exponent.
Close to the transition, the singularity is usually  described by a
{\it power law} or {\it logarithmic} behaviour:
\bea
C_L(\kappa)  &\propto & \vert t \vert ^{-\alpha ^\prime }, \; \; \hbox{for} \; \kappa >
\kappa _c \; , 
\; \;  t := \frac{\kappa_c - \kappa }{\kappa _c} \; ; 
\label{eq:4.5} \\
C_L(\kappa)  &\propto & - \ln \, \vert t \vert \; .
\label{eq:4.5b} 
\ena
We can relate the divergence of $C$ to that of the correlation length
by noting $ \vert t \vert \propto \xi ^{-1/\nu }$. At finite volume
$L$ and for $\kappa = \kappa _c $, the (in the infinite volume
divergent) correlation length $\xi $ is limited by $L$. We then observe
\bea 
C_L(\kappa _c)  & \propto & \left( \vert t \vert ^{-\nu }
\right)^{\alpha/\nu } \; \propto \; \xi ^{\alpha/\nu } \; = \; 
L^{\alpha/\nu } \; , 
\label{eq:4.6} \\ 
C_L(\kappa _c)  & \propto & \ln \xi \; = \; \ln L \; . 
\label{eq:4.6b} 
\ena
The later equations are ideal to study finite size scaling: We
calculate the specific heat for several 
values $L$ as a function of $\kappa $. At the (pseudo-) critical
coupling $\kappa _\mathrm{max} \approx \kappa _c$, the specific heat
is maximal. From the function
$$
L \to C_L(\kappa _\mathrm{max} ) \; , 
$$
we then can infer the scaling via (\ref{eq:4.6}) or (\ref{eq:4.6b}).

\medskip
Figure~\ref{fig:5}, left panel, shows the specific heat as a function
of $\kappa $ for several lattice sizes $L$. We observe the signature
of a 2nd order phase transition: the maximum of the specific heat
increases with increasing $L$, and the position of the maximum, i.e.,
$\kappa _\mathrm{max}$ moves closer to the infinite volume limit
$\kappa _c=1$. Figure~\ref{fig:5}, mid panel, shows the evolution of
$\kappa _\mathrm{max}$ with the inverse system size. We model the
finite size corrections to the critical value by the power series
$$
\kappa _\mathrm{max} \; = \; 1 \; + \; a_1 \, \left(
  \frac{1}{\sqrt{L}} \right) \; + \; a_2 \, \left(
  \frac{1}{\sqrt{L}}\right)^2 \; + \; \ldots  \;  . 
$$
We find that the first three terms of the expansion excellently fit
the curve for $L$ ranging from $50$ to $10,000$.

\medskip
Let us now investigate the scaling with the system size $L$.
Figure~\ref{fig:5}, right panel, indicates that the scaling wth $L$
is logarithmically slow. The data for $L>1000$ are well represented by
the fit function:
$$
C_L(\kappa _\mathrm{max})\; = \; 0.90 \, \ln L \; - \; 0.0027 \, \ln
^3 (L) \; . 
$$
The scaling is therefore similar to the case of 2D Ising model with
standard next-to-nearest neighbour interaction, which is also
proportional to $\ln L$. 

\subsection{Scaling of the correlation length $\xi $ \label{sec:bin}}

\begin{figure*}  
\includegraphics[height=5cm]{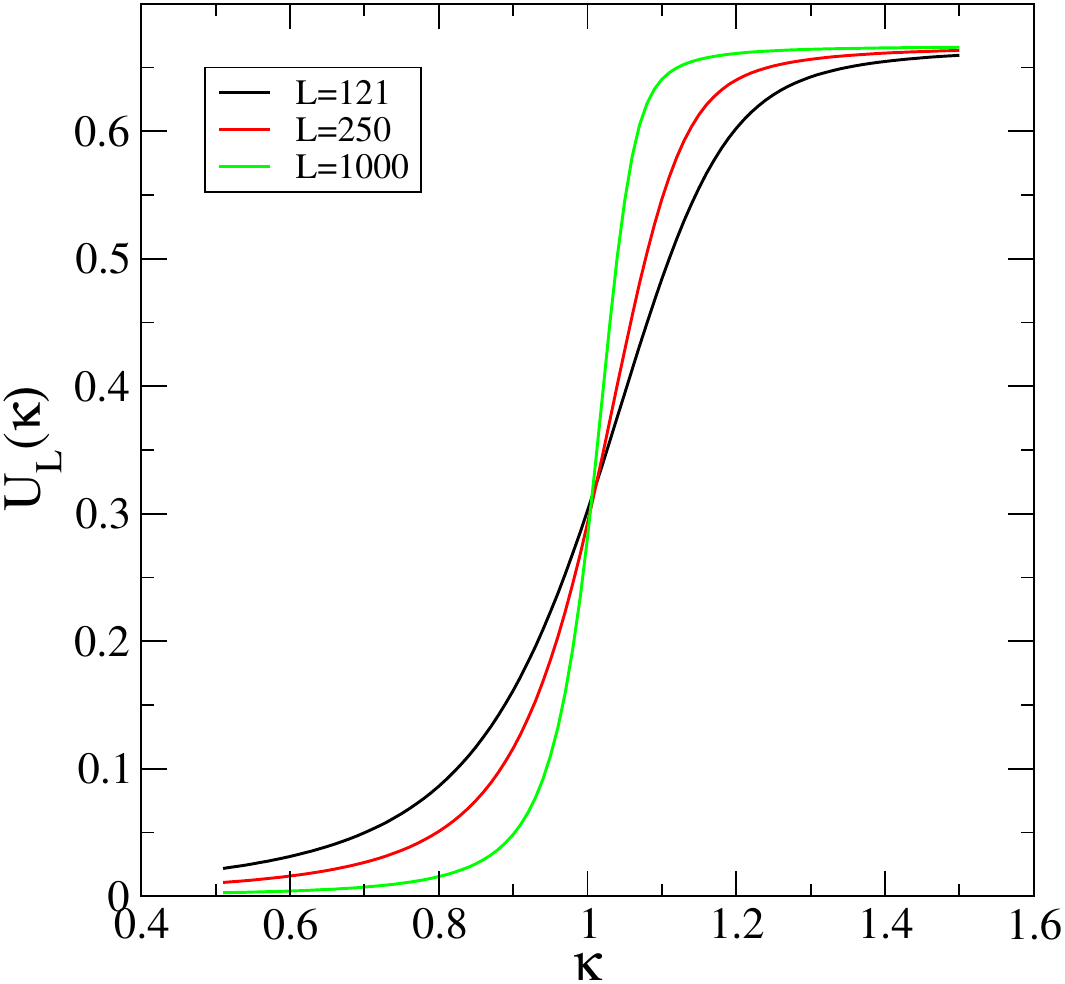}  \hspace{0.5cm}
\includegraphics[height=5cm]{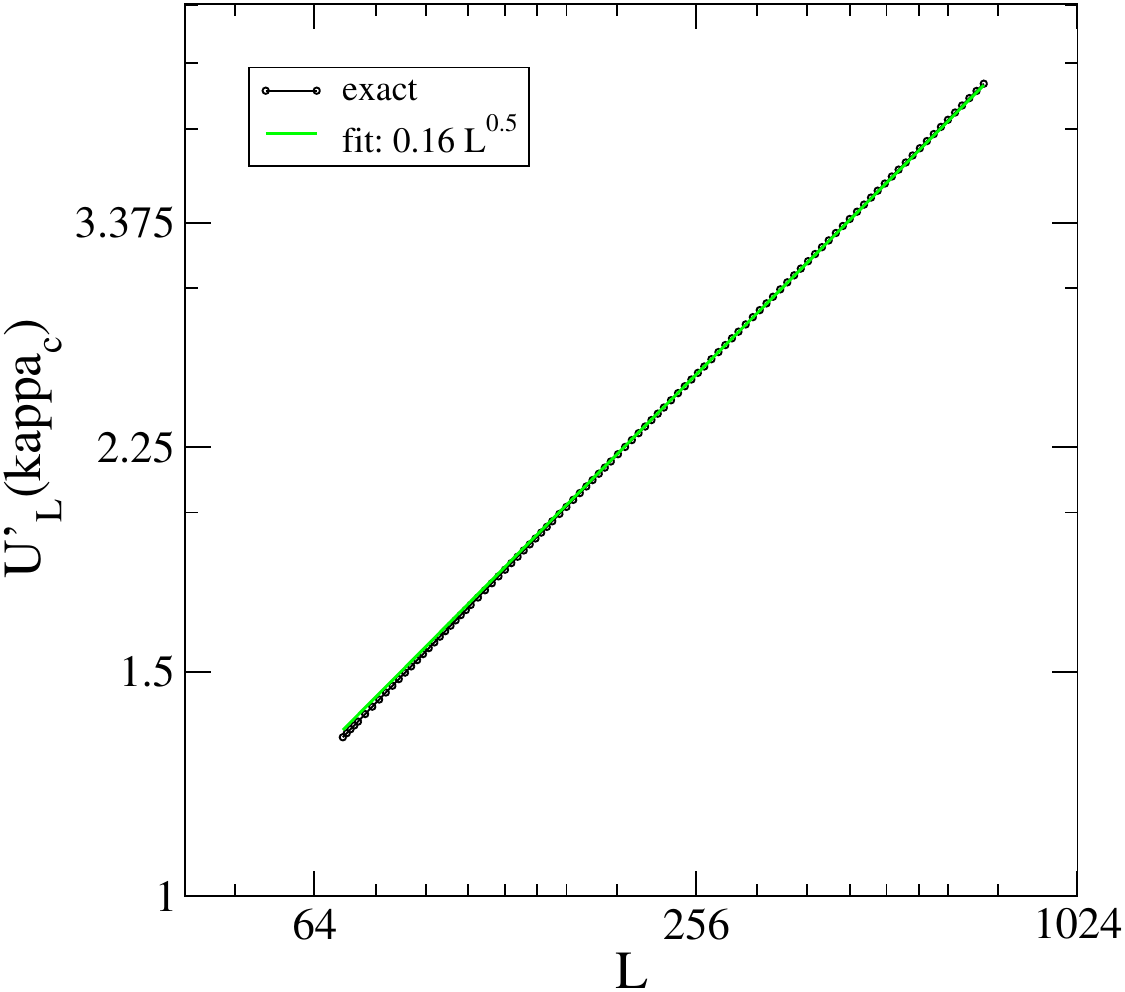}  \hspace{0.5cm}
\includegraphics[height=5cm]{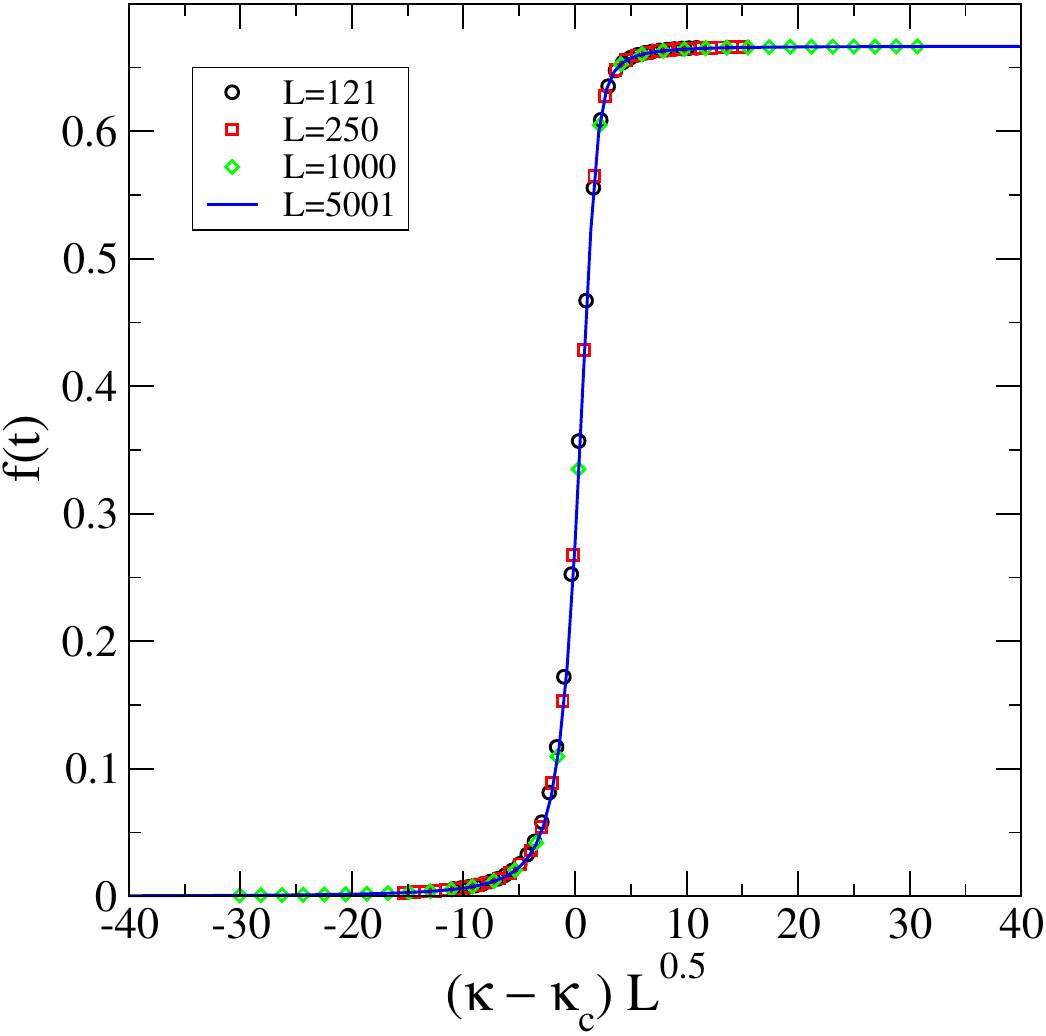}  
\caption{\label{fig:6} Left: The Binder cumulant (\ref{eq:4.11})
  as a function of $\kappa $ for three system sizes $L$. Middle: The
  derivative of the Binder cumulant in a log-log plot revealing the
  scaling (\ref{eq:4.13}). Right: Finite size scaling function $f$ for
  three values of $L$. 
}
\end{figure*}

A convenient way to
extract the critical coupling and the exponent $\nu $ is by means of the
Binder cumulant:
\be
U_L \; = \; 1 \; - \; \frac{ \langle O ^4 \rangle }{ 3 \, \langle O^2
  \rangle ^2 } \; ,
\label{eq:4.11}
\en
where $O$ is the order parameter of the statistical system. In the
broken phase, order parameter expectation values are dominated by a
non-vanishing scale $O_s$:
$$
\langle O \rangle =0 \; \; \; \hbox{(finite L)} , \hbo
\langle O^4 \rangle \approx O_s^4, \hbo 
\langle O^2 \rangle \approx O_s^2, 
$$
which suggests that the Binder cumulant approaches $2/3$ for large
couplings $\kappa \gg \kappa _c$. In the disordered phase, the
probability distribution of the order parameter $O$ is approximately
Gaussian: the moments fulfil the relation
$$
\langle O^4\rangle \; \approx \; 3  \, \langle O^2 \rangle ^2
$$
and the cumulant vanishes for $\kappa \ll \kappa _c$.

\medskip
At the critical coupling $\kappa = \kappa _c$, the Binder cumulant
$U_L$ becomes independent of the system size (to leading order $1/L$),
leading to a crossing point when $U_L$ is plotted as a function of
$\kappa $ for several system sizes $L$. The critical value $U_L$ for
$\kappa = \kappa _c$ depends on the dimensionality of the system and
the symmetry of the order parameter, but it is otherwise universal. It
therefore can be used to categorise  statistical systems in
universality classes.

\medskip
Close to the phase transition, the only relevant ratio of scales is the
system size $L$ over the correlation length $\xi $. The Binder cumulant
then obeys a finite-size scaling form: 
\be
U_L(\kappa ) \; = \; f \left( (\kappa - \kappa _c ) \,
  L^{1/\nu}   \right)  \; ,
\label{eq:4.12}
\en
where $f$ is a universal scaling function.

\medskip
The critical exponent $\nu $ is conveniently estimated from a a study
of the derivative of the Binder cumulant:
\be
\lim _{\kappa \to \kappa _c^+} \frac{d}{d\kappa } \, U_L(\kappa ) \; =
  \; L^{1/\nu } \; f^\prime(0^+) \; . 
\label{eq:4.13}
\en
With (\ref{eq:4.11}), we find:
\bea
\frac{d}{d\kappa } \, \langle O^n \rangle &=&
\langle O^n A_2 \rangle \; - \; \langle O^n \rangle \langle A_2
\rangle ,
\nonumber \\ 
\frac{d}{d\kappa } \, U_L(\kappa ) &=& - \frac{ \langle O^4 A_2
  \rangle - \langle O^4 \rangle \langle A_2  \rangle }{3 \, \langle
  O^2 \rangle ^2 }
\nonumber \\
&+& \frac{2  \langle  O^4 \rangle [ \langle  O^2 A_2\rangle -  \langle
  O^2 \rangle \langle A_2  \rangle ]  }{ 3 \, \langle  O^2 \rangle ^3 } \; .
\label{eq:4.14}
\ena

\medskip
Before we can implement this programme, we still need to settle the
question of the order parameter $O$. The $S^2$ theory has a
ferromagnetic anti-ferromagnetic symmetry. The next-to-nearest
neighbour interaction $S$ changes sign under this symmetry
transformation, since it is positive in the ferromagnetic phase and
negative in the anti-ferromagnetic one. We therefore choose
\be
O \; = \; S[z]/(L-1)
\label{eq:4.15}
\en
Again all results can be obtained exactly, and the one for the Binder
cumulant is shown in figure~\ref{fig:6}, left panel. We find that the
cumulants for three sizes $L$ intersect close to the critical coupling
$\kappa _c=1$. Asymptotically, the intersection point approaches the
critical coupling but the approach is slow due to the ``square-root''
behaviour and thus large finite size corrections (see figure~\ref{fig:5},
middle). We have also calculated the derivative of the Binder cumulant
(\ref{eq:4.14}) at the critical coupling $\kappa _c$ as a function of $L$
(see figure~\ref{fig:6}, middle). The result in the log-log plot is
perfectly fitted by a straight line bearing witness to the scaling law
(\ref{eq:4.13}). From the fit, we find the critical exponent is 
compatible with 
\be
\nu \; \approx \; 2 \, .
\label{eq:4.16}
\en
Figure~\ref{fig:6}, right panel, shows the finite size scaling
function $f$ in (\ref{eq:4.12}) as a function of the scaling variable
$$
t := (\kappa - \kappa_c) \, L^{1/\nu} \; , \hbo \nu =2 . 
$$
We perform a scaling test by extracting $f$ for three different 
lattice sizes. We find excellent scaling over a wide range of $t$. Extracting the critical Binder value $B_c \; = \; f(0) $ from the
calculation with the largest system size $L=5001$, we find
\be
B_c \; \approx \; 0.2753...
\label{eq:4.17}
\en
For a horizon screening we list the critical exponent $\nu $ and the
Binder Cumulant at criticality for several well studied models:

\medskip
\begin{tabular}{cclll}
    model & dim & $\nu $ & $B_c$ & comment \\ \hline \hline
    Ising & 2D & 1 & 0.6107.. & exact solution \\
    Ising & 3D & 0.630.. & 0.465...  & numerical \\
    O(2) & 2D & BKT & 0.636... & no true criticality \\
    O(2) & 3D & 0.671.. & 0.586 & superfluid helium \\
    O(3) & 3D & 0.711.. & 0.618.. & rotational sym break \\
    3-state Potts & 2D & 5/6 & 0.613.. & 2nd order \\
    4-state Potts & 2D & 2/3 & 0.66.. & log corrections \\
    Ising & $\ge \, $4D & 1/2 & 0.2705 & Landau theory applies \\
  \hline \hline 
\end{tabular}

\medskip
When it comes to the 1D-Ising model with a long range interaction of
the type $J(r) \propto 1/r^{1+\sigma}$, $\sigma < 1$, the model
possesses a phase transition for $0 < \sigma  \le 1/2$ 
with mean field critical exponents. For $1/2  < \sigma  < 1$, both
$\nu $ and $B_c$, vary with $\sigma $~\cite{Benedetti2025}.  Higher 
dimensional field theories with local interactions have typically $\nu
<1 $. 

\section{Generalisation to continuous symmetries \label{sec:o3}}

\subsection{The 1D $O(3)$ model with $S^2$ action } 

Continuous symmetries play an important for the inhibition of spontaneous symmetry breaking in low dimensions. The reason is the existence of massless excitations - the Goldstone boson - that creates infrared singularities in low dimensional momentum integrals. The role of the Goldstone boson can only be explained beyond perturbation theory and is to effectively restore the broken symmetry (Coleman theorem~\cite{Coleman1973}). It is therefore important that the above outlined mechanism is still in operation for low dimensional field theories with continuous symmetry. 

\medskip
We study here a generalised $O(3)$ model with unit vectors $\vec{n} \in \R^3$ as the degrees of freedom. The order parameter is given by 
\be 
S[\vec{n}] \; = \; \sum _{l \in \langle xy \rangle } \vec{n}_x \cdot \vec{n}_y \; . 
\label{eq:5.1}
\en
For the field dependent mesoscopic interaction, we choose one that reinforces the (anti-)ferromagnetic interaction. The interaction strength and partition function are given by  
\bea 
\beta \Bigl( S[\vec{n}] \Bigr) &=& \frac{ \kappa }{ 2 \, (L-1) } \; S [\vec{n}] \; , 
\label{eq:5.2} \\
Z(\kappa ) &=& \int \prod _ x [d \Omega _x] \; \exp \left\{ \beta \Bigl( S[\vec{n}] \Bigr) \; 
S[\vec{n}] \right\} \; ,  
\label{eq:5.3}
\ena 
where $\Omega _x $ is the angular parametrisation of the unit vector and $d \Omega _x$ is the related and $O(3)$ invariant Haar measure: 
$$
\vec{n} \; = \; \left( \begin{array}{c} 
\sin \theta \, \cos \varphi \\
\sin \theta \, \sin \varphi \\
\cos \theta 
\end{array} \right) \; , \hbo d \Omega \; = \; \, d\varphi \, d\theta \, \sin \theta \; . 
$$
We closely follow the calculation of the partition function for the $S^3$-Ising model in subsection~\ref{sec:IsingS2} and re-write the partition function in terms of the density of states $\rho (E)$: 
\bea
Z (\kappa ) &=& \int dE \;  D^{(2)} (E) \; ,
\nonumber \\
D^{(2)} (E) &=& \rho(E) \; \exp \left\{ \frac{ \kappa }{2 (L-1)} \, E^2
  \right\} \; , 
\label{eq:5.4} \\
\rho (E) &=& \int \prod _ x [d \Omega _x] \; \delta \left( E \; - \; \sum _{l \in \langle xy \rangle } \vec{n}_x \cdot \vec{n}_y \right) \; . 
\label{eq:5.5} 
\ena
We point out the similarity of Ising and $O(3)$ theory in this formulation. Differences only manifest themselves in the density of states. We are going to estimate the density of states 
and will argue that the qualitative features of $\rho $ are the same so the conclusion of the existence of a phase transition is unchanged. 

\subsection{Calculation of the density of states} 
\begin{figure*}  
\includegraphics[height=5.2cm]{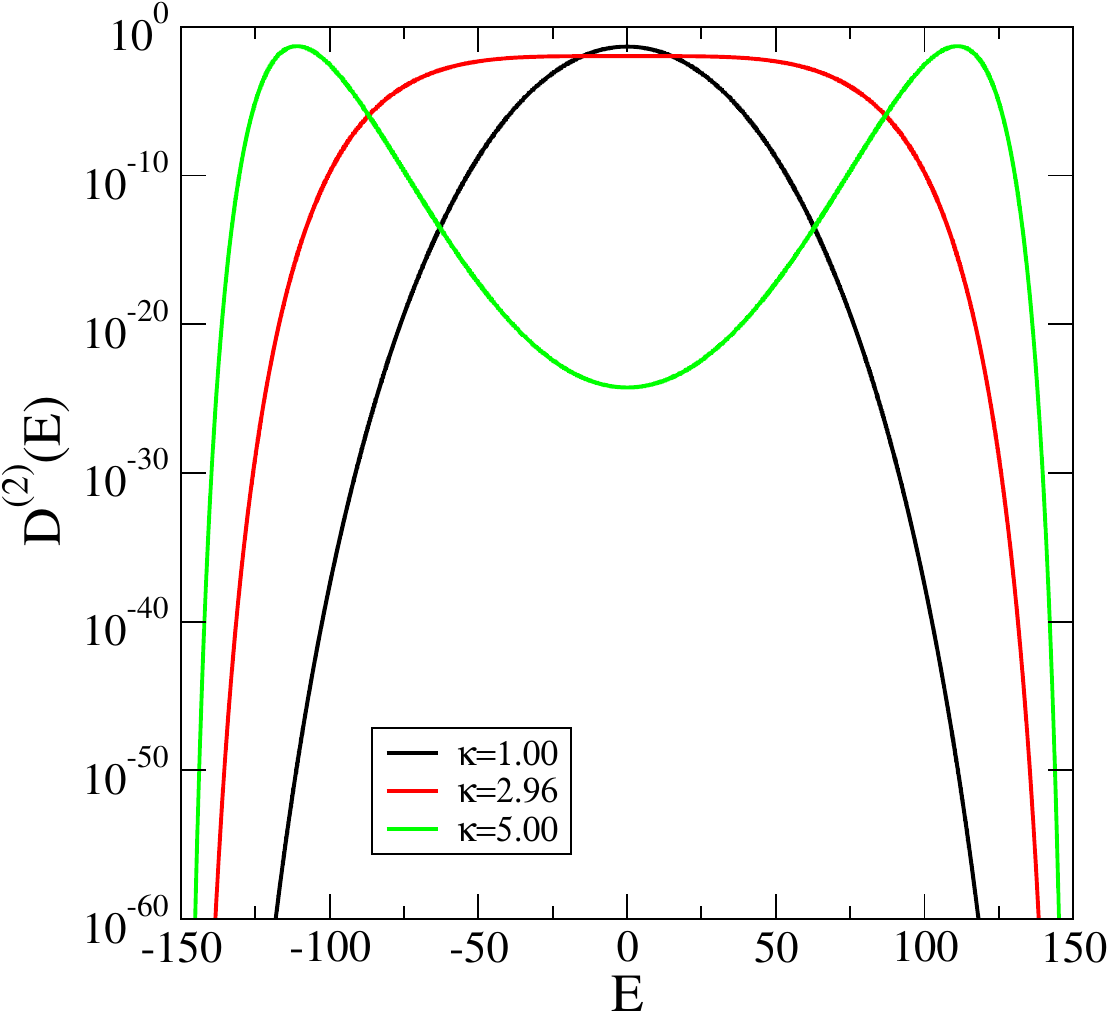}  \hspace{0.5cm}
\includegraphics[height=5.2cm]{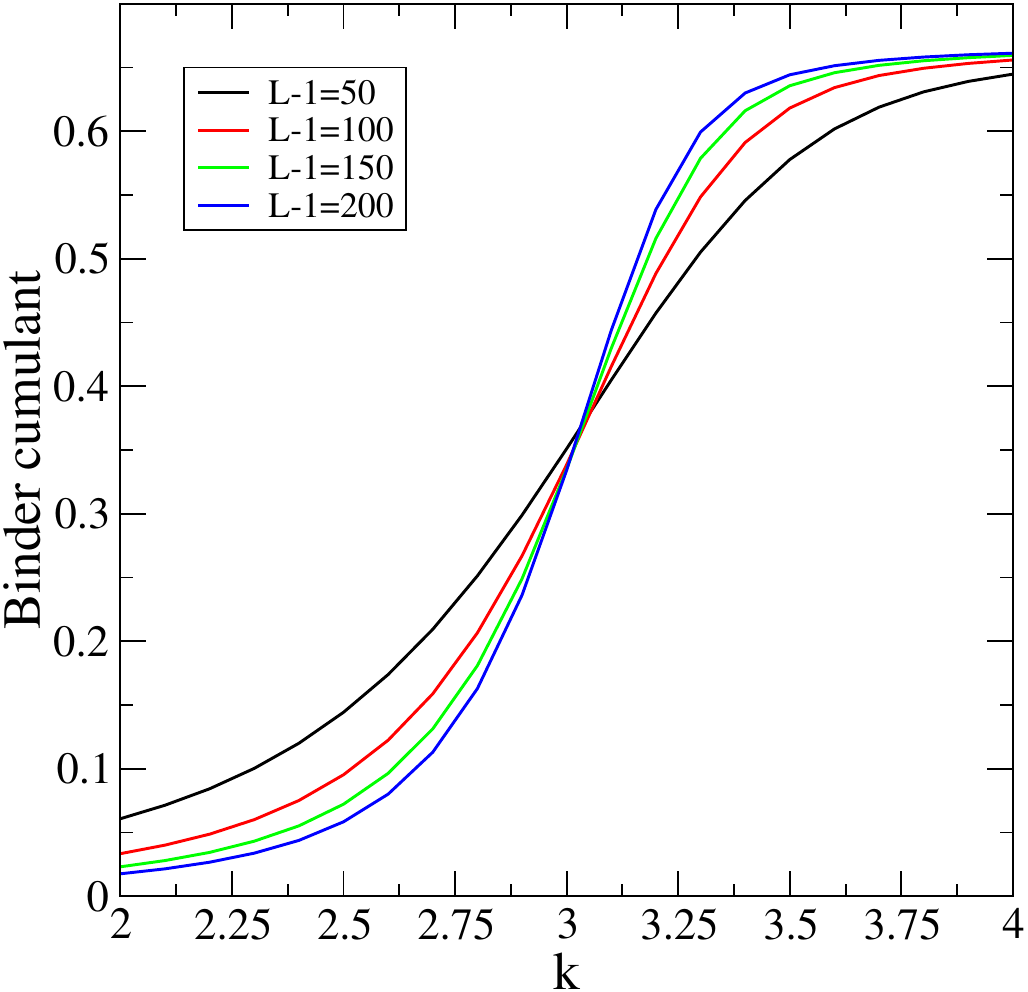} \hspace{0.5cm}
\includegraphics[height=5.2cm]{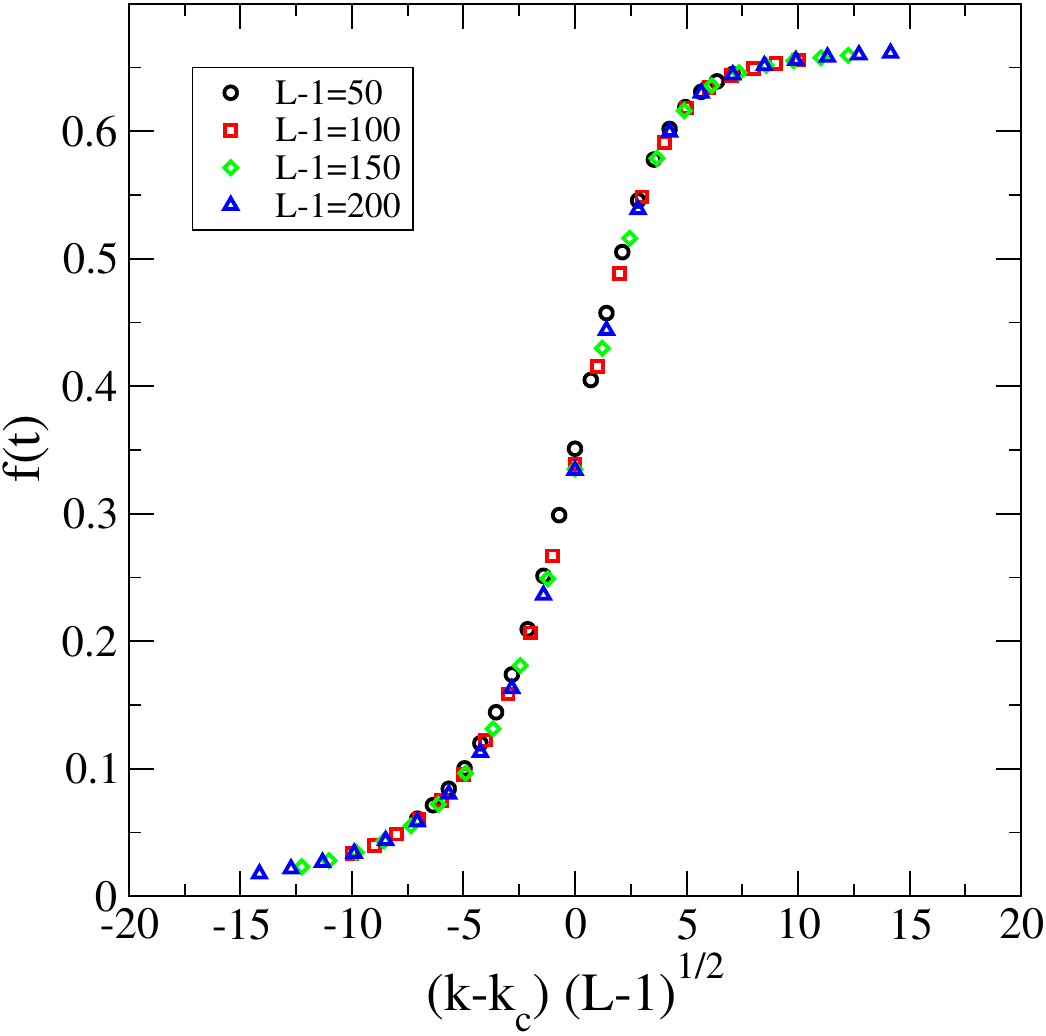}
\caption{\label{fig:7} {\bf 1 dimensional O(3) model.} Left: the marginal distribution  $D^{(2)}(E)$  function of the action $E$ for $151$ spins. Middle: The Binder cumulant.  Right: The scaling function with a critical index $\nu \approx 2$. }
\end{figure*}

The big technical advantage of the 1D theory over those in higher dimensions is that the  theory can be transformed into a non-interaction theory by a change of degrees of freedom. Consider the unit vector at position $z>1$, i.e., $\vec{n}_z$, with angle $\theta _z$. We can always perform a $O(3)$ rotation in the $z$-coordinate system that brings the unit vector $\vec{n}_{z-1}$ to the $z$-axis. This transforms the interaction: 
$$ 
\cos \Bigl( \theta _{z-1} - \theta _z \Bigr)  \; = \; \vec{n}_{z-1} \vec{n}_{z}  \; \to \; 
\vec{e}_z \vec{n}_{z} \; = \; \cos \theta _z \; . 
$$
The Haar measure $d\Omega _z$ is invariant under this transformation.  We perform a further variable transformation: 
$$ 
d \varphi \; d \theta \; \sin \theta \; = \; d \varphi \; d \eta \; , \hbo \eta \in [-1,1] \; . 
$$
We point out that the integrands in (\ref{eq:5.5}) do not depend on the angles $\varphi $ and neither on $\theta _1$. Integrating those variables produces a constant multiplicative factor. Altogether we find: 
\bea 
\rho (E) & = &  2 (2\pi )^{L-1} \, \int _{-1}^1 \prod_x d \eta _x 
\nonumber \\ 
&& \delta \Bigl( E \; - \; \left[
\eta _1 + \ldots + \eta _{L-1} \right] \Bigr). 
\label{eq:5.6} 
\ena 
Since $\-1 \le \eta \le 1$, we find that the density of states is only non-vanishing for the allowable energy range, i.e., 
\be 
\rho (E) = 0 \; , \; \; \hbox{for} \;  E > L-1 \; \hbox{ or for}\; E <- (L-1) \; . 
\label{eq:5.6b} 
\en
For studying the density of states in the large particle number limit and moderate energies, 
 the Fourier representation of the Dirac $\Delta $-function provides a convenient starting point. We  calculate:   
\bea 
\rho (E) & = &  2 (2\pi )^{L-1} \, \int dx \; \e^{ixE} \; \int_{-1}^1 \prod_k d \eta _k \;  
\nonumber \\ 
&& \exp \Bigl\{- ix [\eta _1 + \ldots  +  \eta _{L-1} ] \Bigr\} 
\nonumber \\ 
& = &  2 (2\pi )^{L-1} \, \int dx \; \e^{ixE} \; \left( \int_{-1}^1 d \eta \; \e^{-ix\eta } \right)^{L-1}
\nonumber \\ 
& = &  2 (4\pi )^{L-1} \, \int dx \; \e^{ixE} \; \left( \frac{\sin x}{x} \right)^{L-1}
\label{eq:5.7} 
\ena
For large $L$, the $(\sin x /x)^{L-1} $ function sharply peaks at $x=0$ with other maxima strongly suppressed. We therefore adopt a semi-classical approximation: 
\bea 
\rho (E) & \approx & 2 (4\pi )^{L-1} \, \int dx \; \e^{ixE} \; \exp \Bigl\{ - \frac{(L-1)}{6} \, x^2 \Bigr\} 
\nonumber \\
&=& 2 (4\pi )^{L-1} \, \sqrt{\frac{6 \pi }{L-1}} \; \exp \Bigl\{ - \frac{3}{2 \, (L-1)} \, E^2 \Bigr\} 
\; . 
\label{eq:5.8} 
\ena 
The LLR coefficient (\ref{eq:2.10}) as a function of $e = E/(L-1)$ 
is given by: 
\be 
\bar{a}(e) \; \approx \;  3 \, e \; , \hbo \hbox{ for} \; e \ll 1\; . 
\label{eq:5.9} 
\en
The limit $e \to 1 $ is beyond the semi-classical approximation: we already know that $(e \to 1) \to \infty $ since the density of state vanishes at the highest possible energy. 
The equation for extrema (\ref{eq:2.12b}) of the $D^(2)(E)$ is then given by: 
\be 
\bar{a}(e) \; = \; \kappa \,e \; \; \; \Rightarrow \; \;\;   3 \, e \; \approx \; \kappa \, e \; \; \; 
(e \ll 1) \; . 
\label{eq:6} 
\en
We conclude that for $\kappa < 3 $, the only extremum (maximum) is at $E=0$: the system is in the disordered phase. For $\kappa >3 $, the right-hand side rises more quickly than the left hand side. Since $a(e \to 1)$ diverges, there must be another intersection point at $e>0$ (and its mirror image at $e<0$) in addition to the extremum at $e=0$. This implies that $e=0$ becomes a minimum, and the extrema at $\pm E_0$ correspond to maxima. 

\medskip
We have calculated the critical coupling $\kappa _c$ in the limit of large $L$. We now will embark  calculating the density of states in analytical form for finite $L$. For any $n$ with $1< n \le L-1$, we define $\rho _n$ and observe the recursion: 
\bea
\rho _n(E) &=& 2 (2\pi) ^n \int _{-1}^{1} \prod _{x=1}^n d \eta _x \; 
\delta \Bigl( E - \eta _1 - \ldots - \eta _{n} \Bigr)
\nonumber \\ 
&=& 2 \pi \int _{-1}^1 d \eta \; \rho _{n-1} \Bigl( E - \eta \Bigr) \; . 
\label{eq:6.2} 
\ena
The desired density of states is then given by 
$
\rho (E) = \rho _{L-1}(E) \; , 
$
and the recursion's starting point is: 
\be
\rho _1 (E) \; = \; \left\{ \begin{array}{ll} 
4 \pi  & \hbox{for} \; \; E \in [-1,1] \\ 
0 & \hbox{else.} 
\end{array} \right. 
\label{eq:6.3} 
\en
We can find the density of states by means of $L-1$ integrals. In practice, this involves a representation of $\rho $ in function space and an approximation of the integrals (numerical integration). For this first study, we have represented $\rho $ as equally spaced points on a line between $-(L-1) \ldots (L-1)$. We found that an approach with $300 x (L-1)$ points yields stable results. Details will be presented in future work. Note we have normalised the densities $D^{(2)}(E)$ in (\ref{eq:5.4}) by imposing: 
$$ 
\int dE \; D^{(2)}(E) \; = \; 1 . 
$$
We find that, due to a finite $L=151$, the critical coupling $\kappa _c= 2.96(1)$ is slightly smaller than the infinite $L$ result of $3$. 
The results are then shown in figure~\ref{fig:7}, left panel, for $\kappa =1$ (disordered phase), $\kappa = 2.96$ (at criticality) and $\kappa =5$ (ordered phase). Note that in the broken phase, the symmetric point $E=0$ is suppressed by about $20$ orders of magnitude.

\subsection{Spontaneous Symmetry Breaking in the $S^2$ O(3) spin chain }

For any finite number of spins $L$, the O(3) symmetry remains unbroken: the expectation value $\langle S \rangle $ receive equal contributions with opposite sign from the relevant regions around $E_0$ (ferromagnetic order, i.e. $S>0$) and  $-E_0$ (ant-ferromagnetic order $S<0$). For $L \to \infty $, the tunnelling probability from relevant states with $ E\approx E_0$ to those with $E \approx - E_0$ vanishes: the system "freezes" with either ferromagnetic or anti-ferromagnetic order - the O(3) symmetry is spontaneously broken to a residual $Z_2$ symmetry in either case. 

\medskip
We have then studied the Binder cumulant using the approach from subsection~\ref{sec:bin}. The unscaled cumulants are shown in figure~\ref{fig:7} (middle) as a function of $\kappa $ for the four system sizes  $L = 50, \, 100, \, 150 \, 200 $. The intersection of all four curves at the critical coupling $\kappa _c = 3$ is satisfactorily with some small scaling violations for the $L=50$ case. The universal intersection point is compatible with $1/3$, i.e., 
\be 
B(\kappa _c) \; = \; 0.33(1) \; .
\label{eq:6.4} 
\en
We have then extracted the critical exponent $\nu $ using (\ref{eq:4.13}) and (\ref{eq:4.14}) and find that 
\be 
\nu  \; \approx \; 2 . 
\label{eq:6.5} 
\en
Equipped with this result and using $\kappa _c=3$, the finite size scaling function for the rescaled Binder cumulant is shown in figure~\ref{fig:7}, left panel, which displays satisfactorily scaling for all system sizes involved. 

\medskip
This model deserves further studies, which are left to future work: the scaling properties of the magnetic susceptibility in relation to the correlation length is an interesting case study to show universality in 1 dimension. Also the fate of the Goldstone boson, the propagation of which induces infrared singularities at least in perturbation theory (Feynman graphs), is also worthwhile to explore.

\section{Conclusions \label{sec:conclusions}}

We have constructed a novel family of one‐dimensional field theories
in which coupling strengths depend explicitly on mesoscopic
observables (e.g. magnetisation or energy density), thereby generating
effective infinite‐range interactions that evade the constraints of
the Mermin–Wagner theorem.  As a consequence, these models sustain
bona fide critical phenomena—including spontaneous symmetry breaking
and emergent long‐range order—despite their intrinsically low
dimensionality.   

\medskip 
By augmenting the conventional nearest‐neighbour Ising action with
non‐local feedback terms proportional to powers of the local action
(denoted $S^2$ and $S^3$), we identify two new 1D universality
classes, each characterised by critical behaviour irreducible to any
known class in one, two, or three dimensions.  In particular, we
observe:

\begin{itemize}
\item The $S^2$‐model undergoes a continuous, second‐order phase
  transition at $\kappa_c=1$.  The specific heat diverges logarithmically
  with system size and the correlation‐length exponent is anomalously
  large ($\nu \approx 2$), signalling non‐trivial scaling beyond traditional Ising
  limits.

 \item The $S^3$‐model exhibits a discontinuous, first‐order
   transition at $\omega \approx 2$, marked by an abrupt jump in internal energy
   and the emergence of a dominant ferromagnetic phase.  This
   dichotomy underscores the decisive role of the feedback functional
   form in setting the transition order. 
   
\end{itemize}
\hfill \break 
We have extended the scope of the investigation to the 1-dimensional O(3) model to address the important case of a continuous symmetry. Spontaneous symmetry breaking is accompanied by the emergence of massless Goldstone modes, which could restore the symmetry in low-dimensional field theory. For the case of an $S^2$ mesoscopic motivated interaction, we however find a second-order phase transition at the critical coupling $\kappa _c = 3$. The high-$\kappa $ emerging phase spontaneously breaks the O(3) symmetry, and the emerging phases exhibit either ferromagnetic or anti-ferromagnetic order, both with a residual $Z_2$ symmetry. 

\medskip 
We have further demonstrated that these non‐local interactions admit
natural physical realisations—ranging from the stiffness‐dependent
conformations of dense polymer chains to one‐dimensional effective
theories obtained via dimensional reduction of higher‐dimensional
field models.  Such derivations confirm that mesoscopic feedback and
attendant long‐range couplings can arise organically in both physical
and biological contexts, thereby enhancing the models’ empirical
relevance.

\medskip 
Our analysis is supported by exact analytic calculations of partition
functions, density of states, and Binder cumulants, complemented by
high‐precision numerical studies employing finite‐size scaling to
extract critical exponents.  All results cohere to validate the
presence of genuine phase transitions in one dimension and to
substantiate the classification of these systems into hitherto
unexplored universality classes.   

\medskip 
In summary, one‐dimensional field theories endowed with mesoscopic
feedback–induced infinite‐range interactions not only contravene the
traditional prohibition of 1D criticality but also inaugurate new
avenues for theoretical and applied research.  These findings invite
further exploration of mesoscopically regulated interactions in
complex systems, with potential implications spanning statistical
field theory, soft‐matter physics, and beyond.

\bigskip
\textbf{Acknowledgement: } We are grateful to Slava Rychkov for
helpful comments on the manuscript.  Amanda Turner acknowledges partial support
by EPSRC grant EP/T027940. 

\bibliography{ising_1d_v16_lit}

\end{document}